\documentstyle[preprint,aps,psfig]{revtex}

\begin{document}
\draft
\title{Quasi Two-dimensional Transfer of Elastic Waves}

\author{
Nicolas P. Tr\'egour\`es and Bart A. van Tiggelen}

\address{Laboratoire de Physique et Mod\'elisation des Milieux Condens\'es\\
CNRS/Universit\'{e} Joseph Fourier,
B.P. 166, 38042 Grenoble Cedex 09, France\\
}
\date{\today }
\maketitle

\begin{abstract}
A theory for multiple scattering of elastic waves is presented
in a random medium bounded by two ideal 
free surfaces, whose horizontal size is infinite and whose
 transverse size is smaller than the mean free path of the waves.
 This geometry is relevant for seismic wave propagation in the Earth crust.
We derive a time-dependent, quasi-2D radiative transfer equation,
 that describes the coupling of the eigenmodes of the layer
 (surface Rayleigh waves, SH waves, and Lamb waves).  
 Expressions are found that relate the small-scale fluctuations to
  the life time of the modes and to their coupling rates.
 We discuss a diffusion approximation that simplifies the mathematics
  of this model significantly, and which should apply at large lapse times.
 Finally, coherent backscattering is studied  within the  quasi-2D radiative transfer 
  equation for different source and detection configurations.    
\end{abstract}

\pacs{46.40.Cd, 91.30.Dk, 43.20.Fn, 62.30.+d }

\section{Introduction}

Multiple scattering studies  of elastic waves
have become relevant to get to a deeper understanding of the seismic Coda $Q$ 
\cite{aki1,aki2}
and the principle of equipartition \cite{papa1,hennino}. 
Multiple scattering is believed
to occur in the spectral regime of 1- 15 Hz. Waves with larger frequencies 
suffer from absorption. Waves with smaller frequencies don't see the disorder
any more. 

The  wavelength of a transverse elastic wave at a frequency of 2 Hz is 1.7 km.
This brings us to  a basic problem in seismic studies
 of multiple scattering of elastic waves: 
all measurements take place at the free surface. As a result, they suffer 
from coherent reflections and mode-conversions.  To be of any relevance to seismology,
a multiple scattering theory should be capable of describing the boundaries
in a very precise way, preferentially on the level of the wave 
equation. The standard equation of radiative transfer \cite{chandra}
does not have this property. All phase information has been lost, and
the accuracy of its spatial resolution is estimated to be some small fraction
of the mean free path, estimated to be equal to 20 km
at least for seismic waves which is
usually still much bigger than the wavelength.  The equation of radiative transfer has  
been studied in ultrasonics for elastic waves at much higher frequencies
\cite{hirs,weaver,turnweav1,turnweav2}, for which the 
complication of near-field detection  is much less a problem.  

In this paper we present a multiple-scattering model that has been adapted
to the needs of seismology. It incorporates the complex polarization
properties of elastic waves as well as the mode conversions at the
free surface. At the same time, we can formulate an almost convential 
radiative transfer equation describing mode conversions 
induced by scattering. Many other contemporary phenomena, such as
equipartition and coherent backscattering, can be studied
for measurements taking place near the free surface.

The set-up of this paper is as follows. In section 2 we look at the 
wave equation for elastic waves, and we will define the
Green's function for elastic wave propagation. 
In section 3 we introduce small-scale fluctuations
and define the ensemble-averaged Green's function. This   provides
us with the extinction times of all elastic modes. They will serve 
to define our quasi-2D  approximation. In section 4, the transport equation is
derived, which describes the time-evolution of the ensemble-averaged
energy-contents of all individual modes, and whose stationary solution
exhibits equipartition of energy between all modes. In section 5,
we discuss the application of the well-known diffusion approximation
to this quasi-2D model, introducing a $N \times N$ diffusion tensor
for $N$ modes. Finally, in section 6 we investigate 
coherent backscattering using our Quasi-2D approximation for different 
source and detection configurations. Section 7
is devoted to conclusions and perspectives.

\section{Time-evolution of Elastic Waves}

In this section we will formulate the mathematics of elastic wave
propagation in a way that is suited to apply conventional
methods in wave transport. Many elements have already been discussed very thoroughly 
by Papanicolaou et al. \cite{papa1}, and some will be recalled here for 
convenience.
We start out with   Newton's
second law for the elastic displacement ${\mathbf u}$ at time
$t$ and position ${\mathbf r}$,
\begin{equation}
\rho({\mathbf r})\partial_t^2   u_i = \partial_j\sigma_{ij}({\mathbf r}) + 
  f_i({\mathbf r},t)
\, . \label{n2}
\end{equation}
Here, $\rho({\mathbf r})$ is the local mass density, ${\mathbf f}({\mathbf r},t)$
is an external force per unit volume, and 
$\sigma_{ij}({\mathbf r})$ is the local traction which, by Hooke's law,
is given by \cite{akibook,seis2,MF}
\begin{eqnarray}
\sigma_{ij}({\mathbf r}) &=& C_{ijkl}({\mathbf r}) \varepsilon_{kl}({\mathbf r})
\nonumber \\
&=& \lambda({\mathbf r}) \varepsilon_{kk} \delta_{ij} + 2\mu({\mathbf r})
\varepsilon_{ij}({\mathbf r})
\, , 
\label{hooke}
\end{eqnarray}
with $\varepsilon_{kl} = \frac12 (\partial_k u_l + \partial_l u_k)$ the
stress tensor.
As always, summation over repeated indices is assumed implicitly.
The second equality applies to an isotropic elastic medium, in which case
the four-rank tensor $C_{ijkl}$ can only have two
independent contributions, proportional to the Lam\'e moduli  $\lambda$ and $\mu$. 
Following Papanicolaou et al. \cite{papa1} we shall introduce the  vector field,
\begin{equation}
{\mathbf \Psi}({\mathbf r},t)=
\left ( \begin{array}{c}
\sqrt{\frac{\lambda}{2}}{\mathbf p} \cdot {\mathbf u} \\
\sqrt{\frac{\rho}{2}}i\partial_t u_i \\
-i \sqrt{\mu} \varepsilon_{ij} 
\end{array}
\right ).
\label{psi}
\end{equation}
This vector has 9 independent components since $\varepsilon_{ij}$ is a
symmetric tensor, whose trace is given by the first component. We have introduced the operator ${\mathbf p} = -i \nabla $.
It can readily be checked that Eqs.~(\ref{n2}) and (\ref{hooke})
combine to the following time-evolution problem,
\begin{equation}  
i\partial_t \left| {\mathbf \Psi} ({\mathbf r},t)\right> =  {\mathbf K}({\mathbf r}, {\mathbf p})
\cdot \left|{\mathbf \Psi} (t) \right> + 
\left|{\mathbf \Psi}_f(t) \right>\, ,
\label{te}
\end{equation}
with the time-evolution operator,
\begin{eqnarray}
{\mathbf K} = 
\left(
\begin{array}{ccc} 
0 & \sqrt{\lambda}\vec{{\mathbf  p}}\frac{1}{\sqrt{\rho}} & \vec{\vec{{\mathbf 0}}} \\
\frac{1}{\sqrt{\rho}}{\mathbf p}\downarrow\sqrt{\lambda} & \vec{{\mathbf 0}}\downarrow  &  \frac{1}{\sqrt{\rho}} 
\vec{\vec{{\mathbf L}}}({\mathbf p}) \downarrow   \sqrt{2\mu} \\
{\mathbf 0}\downarrow\downarrow &  \sqrt{2\mu} 
\vec{{\mathbf L}}({\mathbf p}) \downarrow\downarrow   \frac{1}{\sqrt{\rho}}  
& \vec{\vec{{\mathbf 0}}}\downarrow\downarrow   \end{array}
\right)\label{K}
\end{eqnarray}
and the external force-term ${\mathbf \Psi}_f({\mathbf r},t) 
\equiv (0, -{\mathbf f}({\mathbf r},t)/\sqrt{\rho({\mathbf r})},
\vec{\vec{0}} )$. We have introduced the third
rank tensor 
 $  L_{ijk}({\mathbf p}) \equiv \frac{1}{2} \left(p_i\delta_{jk}
+p\delta_{ik} \right)$ and used the formal Dirac notation for
vector fields to facilitate later the more convenient mode base.
The number of arrows determines the order of the tensor. For clarity,
we have put horizontal arrows 
  when they contract in a right-hand side product with a vector. 
 If $\lambda$ and
$\mu$ are real-valued, the matrix ${\mathbf K}$ is manifestly symmetrical
with respect to the ordinary Cartesian scalar product. As a result,
\begin{equation}
\left<{\mathbf \Psi}(t) \left| \right. {\mathbf \Psi} (t) \right> \equiv
\int \mathrm d{\mathbf r} \mathbf\Psi({\mathbf r},t)^*\cdot  
\mathbf\Psi({\mathbf r},t)
= \int  \mathrm d{\mathbf r} \frac{1}{2}\lambda({\mathbf r}) ({\mathbf \nabla} \cdot {\mathbf u})^2 +
\frac{1}{2}\rho ({\mathbf r})(\partial_t {\mathbf u})^2  
+ \mu({\mathbf r}) {\mathrm{Tr}}\, {\mathbf \varepsilon} ^* \cdot {\mathbf \varepsilon}, \label{energy}
\end{equation}
recognized as the total elastic energy \cite{MF}, is conserved in time if
no external forces are present. 
It is customary to split off a term $\mu\, ({\mathrm{curl} }\, {\mathbf u})^2
$ (describing pure shear wave energy) and  $2\mu\, ({\mathrm{div} }\, {\mathbf u})^2
$  (contributing to compressional energy)
from the last term, leaving a rest term $I$. This
identifies four terms as ``kinetic energy", ``compressional
energy", ``shear energy" and an interference term \cite{MF}. The latter
vanishes for plane waves with either pure transverse or pure 
longitudinal polarization.  

Equation (\ref{te}) can easily
be Laplace-transformed (Im $z >0$). This yields the solution
\begin{equation}
\left|{\mathbf \Psi}(z) \right> =   \left[ z-{\mathbf K}\right] ^{-1}  
\left[ 
i\left|{\mathbf \Psi}(t=0) \right> + \left| {\mathbf \Psi}_f(z) \right> \right]
\, .\label{lp}
\end{equation}
The operator $\left[ z-{\mathbf K}\right] ^{-1} \equiv {\mathbf G}(z) $ will
be called the Green's function, and is   introduced here for future need.
It is convenient to define $t=0$ just before the source sets in so that
${\mathbf \Psi}(t=0) =0$ and the force field becomes the source for
  wave propagation.

\section{Extinction of Elastic Waves in a Layer}

We consider a homogeneous elastic plate with mass density $\rho$
 of infinite horizontal dimension
and of thickness  $H$. Both sides of the plate will here be assumed to be   
free surfaces, with traction-free boundary conditions. The Lam\'e 
coefficients are $\lambda$ and $\mu$, in terms of which
the $P$-wave speed is $\alpha\equiv \sqrt{(2\mu+\lambda)/\rho}$ and
the $S$-wave speed is $\beta\equiv \sqrt{\mu/\rho}$.  The eigenmodes
have been discussed in great detail by Weaver \cite{richard1,modes}. Their
representation (\ref{psi}) can be obtained straightforwardly and
we shall denote them by $\{{\mathbf \Psi}_n\}$. The index $n$ is a discrete
index that labels, at constant frequency,  the symmetric and anti-symmetric Lamb and SH waves
in the plate, including the two Rayleigh surface waves (Fig.~\ref{modes}).

The structure of the symmetric and antisymmetric Lamb modes is extremely rich whereas the 
$SH$ modes are ``simple" guided waves. 
For the sake of clarity let us focus on the antisymmetric branches, indicated with normal 
lines in Figure~\ref{modes}. The first antisymmetric mode (first black dot on the right in
Figure~\ref{modes}) is an antisymmetric Rayleigh surface mode.
Its displacement is evanescent for both the compressional and  the shear component.
Rayleigh modes propagate somewhat slower than bulk $S$ or $P$ waves. As a result 
they lie on the right of the two dashed lines indicating 
the pure shear and pure compressional excitations. 
The second antisymmetric Lamb mode (third black dot) lies between the 
two dashed lines indicating 
the pure shear and pure compressional excitations. This mode is evanescent 
for its compressional component but has a propagating shear displacement. It behaves like 
a pure shear mode as we go away from either one of the free surfaces. As a matter of fact its potential energy 
is mostly shear since its compressional potential energy is negligible.
Finally the antisymmetric mode most on the left in Figure \ref{modes} lies on the left of the lines that indicates 
pure shear and pure compressional excitation. Even deep in the plate this mode is a mixture of
$P$ and $S$ displacements. 
As we increase the frequency, the organization of Lamb modes stays intact. 
One finds two surface Rayleigh modes (symmetric and antisymmetric), ``evanescent $P$ but 
bulk $S$" modes and modes that are both bulk $S$ and bulk $P$.

By translational symmetry, the eigenmodes can be chosen proportional to transverse 
plane waves with wave number ${\mathbf k}$. We will treat him initially 
as $\exp(i{\mathbf k}  \cdot {\mathbf x} )/\sqrt{A}$, with  
a discrete contribution of ${\mathbf k} $ to the label $n$
as a result of the periodic boundary conditions
on both sides of a square plate with surface $A$, and eventually take
the limit $A\rightarrow \infty$.

We will now assume the presence of disorder in the plate, to
be specified more precisely later on,
and calculate the   Green's function, averaged over this random disorder.
The exact meaning of this averaging in seismic observations 
will be addressed elsewhere. 
This procedure is the first part to formulate a transport theory \cite{ping}.
Let the disorder be represented by a perturbation $\delta {\mathbf K}$ 
in the time-evolution operator: ${\mathbf K} = {\mathbf K}_0 + \delta {\mathbf K}$. 
The ensemble-averaged ``retarded" (outgoing) Green's function at frequency $\omega$ 
is given by,
\begin{equation}
\left< {\mathbf G}(z=\omega +i0) \right> = \left< {1\over \omega+i0 -{\mathbf K}} 
\right> \equiv  {1\over \omega+i0 -{\mathbf K}_0  -{\mathbf \Sigma}(\omega)}
\,  . \label{dyson}
\end{equation}
This ``Dyson" equation defines the mass-operator ${\mathbf \Sigma}(\omega)$.
The lowest order contribution is given by \cite{frisch},
\begin{equation}
{\mathbf \Sigma} (\omega) = \left< \delta {\mathbf K} \cdot   {1\over \omega+i0 -{\mathbf K}_0} 
\cdot  \delta {\mathbf K} \right>  + {\mathcal O}(\delta {\mathbf K})^3\, .
\label{born}
\end{equation}
Next, we can  insert the complete and orthonormal set $\{{\mathbf \Psi}_n\}$ of the 
homogeneous plate, defined above. Standard first-order perturbation theory yields,
\begin{equation}
{\mathbf G}(\omega) = \sum_n {\left| {\mathbf \Psi}_n \right>   
\left< {\mathbf \Psi}_n\right|
\over \omega -\omega_n -   \Sigma_n(\omega) }\, , \label{dyson2}
\end{equation}
with
\begin{equation}
 \Sigma_n(\omega) =  \sum_m 
\left< \left| \left< {\mathbf \Psi}_n\right| \delta {\mathbf K}
\left| {\mathbf \Psi}_m \right> \right|^2 \right> {1\over \omega   - \omega_m +i0 }
\end{equation}
The imaginary part of this parameter is negative, and is identified with
$-1/2\tau_n$, where $\tau_n$ is the extinction time of mode $n$.

In general, both $\rho({\mathbf r})$, $\lambda({\mathbf r})$ and
$\mu({\mathbf r})$ are random variables. We will  simplify 
the problem by assuming that  $\rho({\mathbf r})$ is constant, and that velocity
fluctuations are due to fluctuations in the two Lam\'e coefficients:
$\lambda({\mathbf r}) = \lambda_0   + \delta\lambda ({\mathbf r})$
and $\mu({\mathbf r}) = \mu_0   + \delta\mu ({\mathbf r})$, with $\lambda_0$
and $\mu_0$ the coefficients of the homogeneous layer. In that case,
\begin{eqnarray}
\delta {\mathbf K} = {1\over \sqrt{\rho}}
\left(
\begin{array}{ccc} 
0 &   (\delta \lambda({\mathbf r})/ 2\sqrt{\lambda_0})\, \vec{{\mathbf p}}\,  
 & \vec{\vec{{\mathbf 0}}} \\
 {\mathbf p} \downarrow \,   (\delta \lambda({\mathbf r}) /2 \sqrt{\lambda_0}) 
 & \vec{{\mathbf 0}} \downarrow  &  
\vec{\vec{{\mathbf L}}}({\mathbf p}) \downarrow   (\delta \mu({\mathbf r})/2\sqrt{2\mu_0}) \\
{\mathbf 0}\downarrow\downarrow &  (\delta \mu({\mathbf r})/2\sqrt{2\mu_0})\, 
\vec{{\mathbf L}} ({\mathbf p}) \downarrow\downarrow    
& \vec{\vec{{\mathbf 0}}}\downarrow\downarrow   \end{array}
\right)\label{dK}
\end{eqnarray}
A straightforward calculation, employing integration by parts, finally leads to,
\begin{eqnarray}
  \left< \left| \left< {\mathbf \Psi}_n\right| \delta {\mathbf K}
\left| {\mathbf \Psi}_m \right> \right|^2 \right> 
&=&  {\omega^2 }\int \mathrm d {\mathbf r}  \int \mathrm d {\mathbf r}'\nonumber \\
& \, & \left\{ \ \   \left< \delta \lambda ({\mathbf r}) \delta \lambda ({\mathbf r}') \right> \, 
(\nabla\cdot {\mathbf u}_n)^*(\nabla\cdot {\mathbf u}_m)
(\nabla'\cdot {\mathbf u}'_n)^*(\nabla'\cdot {\mathbf u}'_m) \right. \nonumber \\
&+&  \left< \delta \mu ({\mathbf r}) \delta \mu ({\mathbf r}') \right>\, 
{\mathrm{Tr}}\, \varepsilon_n^*\cdot \varepsilon_m 
 {\mathrm{Tr}}\, (\varepsilon'_n)^*\cdot \varepsilon'_m \nonumber \\
 &+& \left.  \left< \delta \lambda ({\mathbf r}) \delta \mu ({\mathbf r}') \right>\, 
 (\nabla\cdot {\mathbf u}_n)^*(\nabla\cdot {\mathbf u}_m)
  {\mathrm{Tr}}\, (\varepsilon'_n)^*\cdot \varepsilon'_m + {\mathrm{c.c.}} \ \ 
  \right\}
 \label{alles}
  \end{eqnarray} 
  To evaluate $\Sigma_n(\omega)$ we must specify the 
  spatial correlations between the Lam\'e coefficients. The simplest choice
  is to assume   correlations that are short range 
  with respect to the wavelength,
  \begin{mathletters}\label{correl}
  \begin{eqnarray}
  \left< \delta \lambda ({\mathbf r}) \delta \lambda ({\mathbf r}') \right> 
  &=&   \sigma_\lambda^2(z) \delta({\mathbf r} -{\mathbf r}')\, \\
  \left< \delta \mu ({\mathbf r}) \delta \mu ({\mathbf r}') \right> 
  &=&   \sigma_\mu^2(z) \delta({\mathbf r} -{\mathbf r}')\, \\
  \left< \delta \mu ({\mathbf r}) \delta \lambda ({\mathbf r}') \right> 
  &=&   \sigma_{\mu\lambda}^2(z) \delta({\mathbf r} -{\mathbf r}')\, .
  \label{sr}
  \end{eqnarray}
  \end{mathletters}
 Without extra difficulty, we can still allow for a depth dependence of the 
 correlation functions.
$\Sigma_n$ can now be evaluated for a big plate for which 
$\sum_m \rightarrow \sum_i A \int \mathrm d^2 {\mathbf k}
  /(2\pi)^2$, including a sum over the different branches.
  All factors $A$ cancel if a transverse
 plane wave normalization $\exp(i{\mathbf k}  \cdot {\mathbf x})$ is
 adopted. For the extinction time of mode branch $j$ at frequency
 $\omega$, we find
  \begin{equation}
  {1\over \tau_j(\omega)} = \omega^2   \sum_i n_i \int {\mathrm d^2 \hat{{\mathbf k}}_i  \over 2\pi} 
  \, W(i   {\mathbf k}_i  , j  {\mathbf k}_j) \, .\label{tau}
  \end{equation}
  with $n_i(\omega) \equiv k_i(\omega)/v_i(\omega)$ in terms of the group velocity
   ${\mathbf v}_i = \mathrm d \omega_i /\mathrm d {\mathbf k}_i$.    
 The ``mode scattering cross-section" is defined as,
\begin{eqnarray}
   W(i   {\mathbf k}_i  , j  {\mathbf k}_j )
    &=& \int_0^H \mathrm dz \left\{  \ \ 
  \sigma^2_\lambda(z)
|\nabla\cdot {\mathbf u}_{i {\mathbf k}_i}|^2|\nabla\cdot 
{\mathbf u}_{j  {\mathbf k}_j}|^2 \right. \nonumber 
+  \sigma^2_\mu(z) 
\left|{\mathrm{Tr}}\, \varepsilon_{i {\mathbf k}_i}^*\cdot 
\varepsilon_{j {\mathbf k}_j} \right|^2 \nonumber \\
 &+& \left.  2 \sigma_{\mu\lambda}^2(z) \mathrm{ Re}\, 
 \left(\nabla\cdot {\mathbf u}^*_{i {\mathbf k}_i} \nabla\cdot {\mathbf u}_{j {\mathbf k}_j}
  {\mathrm{Tr}}\, \varepsilon^*_{i {\mathbf k}_i} \cdot \varepsilon_{j {\mathbf k}_j}  \right)
   \ \ \right\}
 \label{ww}
  \end{eqnarray}
  We have chosen to split of the factor $n_i$, so that
  this matrix is symmetric.   
According to our model the extinction time $\tau_j$ does not depend on the direction
of the horizontal wave number ${\mathbf k}_j$.

The {\it imaginary} part of the ensemble-averaged Green's function is directly
related to the excitations of the  waves \cite{pr}. The spectral density
${\mathcal N}(\omega)$ per unit surface can be expressed as,
  \begin{equation}
{\mathcal N}(\omega)  = -\frac{1}{\pi A}   {\mathrm{Tr}}\, 
{\mathrm{Im}}\, {\mathbf G }(\omega) 
= {1\over \pi} \sum_i \int {\mathrm d^2 {\mathbf k}  \over (2\pi)^2}\, 
{1/2\tau_{i {\mathbf k} } \over [\omega -\omega_{i {\mathbf k}} ]^2 + 
1/4\tau_{i{\mathbf k}}^2 } \, . \label{dos} \end{equation}
Due to scattering, all modes are spectrally broadened.  The separation in 
wavenumber  
of two adjacent  modes with the same frequency (see Figure~\ref{modes}) 
is typically of order $1/H$. The uncertainty in $k$ is typically
$1/v_{i{\mathbf k}} \tau_{i{\mathbf k}}$, with $v_{i{\mathbf k}}$ the group
velocity of the mode. 
If 
\begin{equation}
\tau_{i{\mathbf k}} > H/v_{i{\mathbf k}}  \, ,\label{q2d}\end{equation} 
one can assume that different modes  
at fixed ${\mathbf k}$ do not overlap, except at a few degeneration
points where the dispersion curves for modes with different symmetry 
(i.e. SH and Lamb)
cross. This assumption 
is the {\it Quasi 
Two-Dimensional Approximation} (Q2DA). Criterion~(\ref{q2d})
 is typically satisfied in the Earth crust, which has $H\approx 30 $ km,
 a  typical wave speed 
 $\beta \approx 3.5 $ km/s and a mean free time 
 $\tau > 15 $ sec. In the Q2DA we
 find for the spectral density per unit surface  ${\mathcal N}(\omega)= (2\pi)^{-1}
 \sum_i n_i $,
 showing that
$n_i$, defined in Eq.~(\ref{tau}), represents the spectral weight per unit surface of mode $i$ at
 frequency $\omega$ in phase space.

In the following, all time scales will be normalized by the mean free time of 
$S$ waves in an infinite medium with the same amount of disorder. This time  
depends only on $\sigma_\mu^2$ which can be related to the correlation length and the shear velocity 
fluctuations. For a velocity fluctuation of $2\%$ 
and a correlation length of $700m$, both being typical seismic values, we get a shear mean free 
path $\tau_s^\infty \approx 15s$. Note that 
the correlation length is much smaller than the wavelength $\lambda_s=1700m$, which justifies the 
short range approximation of Eqs~(\ref{sr}).

Figure~\ref{extime} shows extinction times for different modes index, 
calculated from Eqs.~(\ref{tau}) and (\ref{ww}),
normalized by the mean free time of $S$-waves in an infinite medium.
 The plate  thickness is
$H = 20.2 \lambda_S$, which has $N= 106$ modes.
The disorder is chosen to be uniform in the whole plate,
and the spatial correlations among the Lam\'e coefficients is taken equal:
$\sigma_\lambda^2=\sigma_\mu^2=\sigma_{\mu\lambda}^2$.
$SH$ modes  show an extinction time very similar to the extinction time of 
$S$-waves in an infinite medium $\tau_s^\infty$. 
On the other hand the Lamb modes present a more complex pattern: 
Rayleigh modes clearly show a shorter 
extinction time, Lamb modes with an evanescent compressional 
component behave very much like
a bulk $S$ wave. Finally, Lamb modes with both bulk compressional
and bulk shear components behave in a complicated fashion but tend to 
have an extinction time larger than $S$ waves in an infinite medium. 
 
In the case of dominant $\mu$ correlation, $\sigma_\lambda^2\ll\sigma_\mu^2$ 
(dominant shear velocity fluctuations) 
the Lamb modes with both ``bulk'' compressional and shear components will have a somewhat relatively larger 
extinction time. On the other hand, 
if the $\lambda$ correlation dominates, $\sigma_\lambda^2\gg\sigma_\mu^2$, 
(strong compressional velocity fluctuations),
the same Lamb modes with `bulk'' compressional and shear displacements will have the shortest
extinction time. 

We would like to point out that the life time of Rayleigh waves is not well described by
our model since they suffer most from surface disorder (fluctuations
in height), which is not included in Eqs.~(\ref{correl}).

\section{Transport Equation in a Layer}

The next task is the formulation of an elastic transport equation in
the Quasi 2D Approximation. Basic observable is the ensemble-averaged intensity
Green's function $\left< {\mathbf G}(\omega^-) \times {\mathbf G}(\omega^+)^* \right>$,
with $\omega^\pm = \omega \pm \frac12 \Omega$. 
It can be expressed in the complete base $\{{\mathbf \Psi}_n \}$ of the homogeneous plate,
giving rise to the matrix element $L(\omega, \Omega)_{nn'mm'}$ (Figure 2). The Bethe-Salpeter 
equation \cite{ping,pr} for this object reads,
\begin{equation}
{\mathcal L}_{nn'mm'}(\omega, \Omega) = G_n(\omega^+)G_{n'}(\omega^-)^* \left[ 
\delta_{nm}\delta_{n'm'}
+ \sum_{ll'} U_{nn'll'}(\omega, \Omega) {\mathcal L}_{ll'mm'}(\omega, \Omega) \right]
\, .\label{bs1}
\end{equation}
with $G_n$ the Dyson Green's function defined in Eq.~(\ref{dyson}), and
a new
object $U$ called the Irreducible Vertex.
Upon introducing $\Delta G_{nn'}(\omega,\Omega) \equiv
G_n(\omega^+) - G^*_{n'}(\omega^-) $ (idem for $\Delta \Sigma$) this equation can be re-arranged
into,
\begin{eqnarray}
\left[ \Omega + (\omega_n -\omega^*_{n'}) - \Delta \Sigma_{nn'}\right]
{\mathcal L}_{nn'mm'}(\omega,\Omega) &=& \Delta G_{nn'}(\omega,\Omega)\Big[
\delta_{nm}\delta_{n'm'}+\nonumber\\
&&\sum_{ll'} U_{nn'll'}(\omega, \Omega) {\mathcal L}_{ll'mm'}(\omega, \Omega) \Big]
\, .\label{bs2}
\end{eqnarray}
This equation is still exact. We will now carry through a number of
  approximations relevant to our problem. 
  For small disorder,
the vertex $U$ is given by \cite{frisch},
\begin{equation}
U_{nn'll'} (\omega, \Omega) = 
\left< \ \    \left< {\mathbf \Psi}_n\right| \delta {\mathbf K}
\left| {\mathbf \Psi}_l \right>  \left< {\mathbf \Psi}_{n'}\right| \delta {\mathbf K}
\left| {\mathbf \Psi}_{l'} \right>  \ \ \right>\, . \label{iv}
\end{equation}
For short-range correlations, as specified in Eqs.~(\ref{correl}),
the vertex $U$ can be straightforwardly related to the cross-section $W(i{\mathbf k}_i,
 j {\mathbf k}_j)$ defined in Eq.~(\ref{ww}).
  For typical wave packets is 
$\Omega \ll \omega$ (i.e. a wave packet contains many cycles) 
so that we neglect $\Omega$ in any functional dependence on frequency 
(``slowly varying envelope approximation").  
The index $n$ consists of one discrete branch index $j$,
and one index ${\mathbf k}$ that becomes  continuous as $A\rightarrow \infty$.
The Q2DA neglects all overlaps between different branches, so that 
$\Delta G(\omega,\Omega)_{nn'} \rightarrow 2\pi i \delta_{jj'} 
\delta[\omega-\omega_j({\mathbf k})]$.   If we let 
${\mathbf k} -{\mathbf k}' ={\mathbf q}$,  
 and    $S_{m}(\omega)$  the source in mode representation,
a new observable quantity $L_{j{\mathbf k}}$ can be defined as
 \begin{equation} 
 \sum_{mm'} {\mathcal L}_{nn'mm'}(\omega,\Omega)S_{m}S^*_{m'} \equiv 2\pi  
 \delta[\omega-\omega_{j{\mathbf k}}] \delta_{jj'}  \times 
 L_{j{\mathbf k}} ({\mathbf q},\Omega). \label{si}
 \end{equation}
In space-time the Q2D transport equation reads,
 \begin{eqnarray}
 \left[ \partial_t + {\mathbf v}_j  \cdot {\mathbf \nabla} +{1\over \tau_{j{\mathbf k}_j}}
 \right] L_{j{\mathbf k}}({\mathbf x}, t)
 &=& \left|S_{j  {\mathbf k}}(\omega)\right|^2 \delta(t) \delta({\mathbf x})\nonumber\\
&& +  \omega^2 \sum_{j'} \int 
 { \mathrm d^2 {\mathbf \hat{k}}_{j'}  \over  2\pi  } \, n_{j'}
  W(j{\mathbf k}_j,
 j'{\mathbf k}_{j'})  L_{j'{\mathbf k}_{j'}} ({\mathbf x},t).\label{finalert}
 \end{eqnarray}
 We will use this equation as a starting point for our calculations.
  The equation is essentially two-dimensional, with a finite
 number of modes (of order $2H\omega/\beta)$ to take care of the third,
 vertical dimension. The great advantage of 
 this equation is that the boundary conditions of the elastic waves
have been dealt with {\it exactly}, i.e. on the level of the wave equation, 
 contrary to conventional transport equations \cite{papa1,turnweav1,turnweav2}.  
 We see that $L_{j{\mathbf k}}({\mathbf x}, t)$ 
 can be interpreted as the {\it specific intensity}
 of the mode $(j{\mathbf k}_j)$ at frequency $\omega$, at horizontal position
 ${\mathbf x}$, at a time $t$ after the release of energy by the source.

The source term $S_{j{\mathbf k}}(\omega)$ is given by,
\begin{equation}
S_{j{\mathbf k}}(\omega) = \left< {\mathbf \Psi}_{j{\mathbf k}} \right| \left.
{\mathbf \Psi}_f \right>  
=  \omega  \int \mathrm d^3 {\mathbf r} \, {\mathbf f}^*({\mathbf r}, \omega)
\cdot {\mathbf u}_{j{\mathbf k}}({\mathbf r})  \, . \label{source}\end{equation}
Since ${\mathbf u}_{j{\mathbf k}}$ is an eigenfunction for which
the energy (\ref{energy}) has been normalized, we see that $\left|
S_{j{\mathbf k}}\right|^2$
has the  dimension of energy. 
Since $(\Omega, {\mathbf q})$-dependence has been neglected in 
the source, it emerges in our transport equation 
as a $\delta(t) \delta({\mathbf r})$ 
in space-time.

\subsection{Dynamics of the Equipartition Process}

Equation (\ref{finalert}) has one very important property that has
been discussed in great detail in  the literature. By recalling
the expression (\ref{tau}) for the extinction time, it follows
immediately that the specific intensity with the property that
 its total mode energy 
\begin{equation}
\int \mathrm d^2 {\mathbf x} \, L_{j{\mathbf k}_j} ({\mathbf x},t)  = 
\mathrm{constant},
\label{ep}
\end{equation}
is independent of the mode-index $j$ and
independent of the horizontal direction of propagation ${\mathbf k}$,
is a stationary solution for $t >0$ of the transport equation.
All solutions converges to this solution regardless the nature and position of the source.
This implies  that finally 
all modes have an equal share in the {\it total} energy 
contents of the plate.
This phenomenon is called {\it equipartition} 
\cite{hennino,papa2,richard2,richard3,turner}, and is believed to be
a fundamental feature of the solution of most transport equations
at large lapse times, provided absorption is absent, or at least small
\cite{ludoabs}. According to our definition (\ref{si}),
the total spectral energy  per unit surface 
in the regime of equipartition is given by,
\begin{eqnarray}
E_\omega(t) &=& \sum_j   
\int {\mathrm{d}^2{\mathbf k} \over (2\pi)^2 }\, \int \mathrm{d}^2{\mathbf x}  \,   
\, L_{j{\mathbf k}}({\mathbf x}, t) \, 2\pi \delta (\omega-\omega_{j{\mathbf k}})\,
. \nonumber \\
&\rightarrow& {\mathrm{constant}} \times \sum_j n_j\, . \label{en2}
\end{eqnarray}
We will introduce the spectral 
 energy density  $E_i({\mathbf x},t)$ of mode $i$ per unit surface,
 and its current density ${\mathbf J}_i({\mathbf x},t)$ according to,
 \begin{mathletters}\label{ec}
  \begin{eqnarray}
  E_i({\mathbf x},t) &\equiv& \int {\mathrm d^2 {\mathbf k} \over (2\pi)^2}
  \, 
  2\pi \delta \left(\omega -\omega_{i{\mathbf k}}\right) \, L_{i{\mathbf k}}
  ({\mathbf x},t) = n_i \int {\mathrm d^2 {\mathbf \hat{k}} \over 2\pi}
  \, L_{i{\mathbf k}_i}({\mathbf x},t)\, , \\
  {\mathbf J}_i ({\mathbf x},t) &\equiv& \int {\mathrm d^2 {\mathbf k} \over (2\pi)^2}
  \, 
  2\pi \delta \left(\omega -\omega_{i{\mathbf k}}\right) \, {\mathbf v}_i 
  L_{i{\mathbf k}}
  ({\mathbf x},t) = n_i \int {\mathrm d^2 {\mathbf \hat{k}} \over 2\pi}
  \, {\mathbf v}_i L_{i{\mathbf k}_i}({\mathbf x},t)\, .
  \end{eqnarray}
  \end{mathletters}

An  exact  equation of continuity can be found from Eq.~(\ref{finalert})
by integrating over ${\mathbf k}_i$,  
\begin{equation}
\partial_t E_i({\mathbf x}, t) + {\mathbf \nabla} \cdot {\mathbf J}_i({\mathbf x}, 
t) = \left[ n_i \int 
{\mathrm d^2 {\mathbf \hat k} \over  2\pi }\, \left|S_{i{\mathbf k}_i}(\omega)
\right|^2 \right] \ \delta({\mathbf x}) \delta(t) 
- \sum_j C_{ij}E_j({\mathbf x}, t)  
 \, , \label{cont} \end{equation}
 with the ``mode-conversion  matrix",
 \begin{eqnarray}
C_{ij} =  {\delta_{ij}\over \tau_i}
- \omega^2 n_i  \int {\mathrm d^2{\mathbf \hat{k}}_j\over 2\pi} \, 
W( i{\mathbf k}_i, j{\mathbf k}_j)
 \, . \label{mcon}
\end{eqnarray}
The mode-conversion matrix
${\mathbf C}$ has an eigenvalue $0$ with (left-hand) eigenvector
$\{n_i\}$, associated with the equipartition. 
The $N-1$ nonzero eigenvalues, which can be called ``Stokes parameters", of the mode conversion matrix
${\mathbf C}$ determine the dynamics of the equipartition process.
It depends on the initial conditions, i.e. how the initial
release of energy is distributed among the different modes,
as described by $S_{i{\mathbf k}}(\omega)$.

Figure \ref{eigenC} shows all  eigenvalues of the matrix ${\mathbf C}$ in the
case of a plate of thickness $ H =  20.2\lambda_S$, for which the number
of modes is $N= 106$. The disorder is uniform in the whole plate
and the spatial correlation among all Lam\'e coefficients is chosen equal:
$\sigma_\lambda^2=\sigma_\mu^2=\sigma_{\mu\lambda}^2$.
The time scale has been normalized to the
mean free time of $S$-waves in an the infinite medium, which has the same
amount of disorder, i.e. as described by Eqs.~(\ref{correl}). 

The largest eigenvalue (associated with the shortest life-time) has  
an eigenvector made of the symmetric and antisymmetric Rayleigh modes.
This configuration is very sensitive to the location of the disorder in the plate.
If the plate does not have any disorder close to the two free surfaces 
(at the length scale of a wavelength) the Rayleigh modes, which have 
a penetration length of the order a wavelength, do not suffer from the 
disorder. As a consequence, their life-time would become very large compared to
the mean free time of $S$-waves in an infinite medium. On the other 
hand, if the disorder is localized close to the free surface the Rayleigh modes
end up with a very large eigenvalue.  
The eigenvectors associated with the flat plateau in Figure~\ref{eigenC} consist of modes whose shear
component strongly dominates over the compressional part. 
As a result their eigenvalues 
are very similar to the inverse shear mean free time of a $S$ wave in an infinite medium. 
The eigenvectors associated with the eigenvalues smaller than unity exhibit a strong
compressional component. They are associated with longer life times as shown 
in Figure~\ref{extime}. Quite logically they show up with a smaller eigenvalue 
(associated to a longer life-time) in the mode-conversion matrix ${\mathbf C}$

In the case of dominant $\mu$ correlation, $\sigma_\lambda^2\ll\sigma_\mu^2$, the picture
does not change drastically since Lamb modes are always dominated by shear. 
For dominant $\lambda$ correlation, $\sigma_\lambda^2\gg\sigma_\mu^2$,
the structure of eigenvalues of the  mode-conversion matrix 
${\mathbf C}$ is modified considerably. 
Eigenvalues that were previously associated with ``bulk" $P$ and $S$ 
vectors now see their life-time becoming much shorter. An eigenvector  with a small 
eigenvalue in Figure~\ref{eigenC}, achieves
a large eigenvalue.

Figures~\ref{EnergyModes} show, for different 
kind of sources, how the initial
release of energy is distributed among the different modes. Figure~\ref{EnergyModes}$a$ shows an isotropic
explosion at a depth $\lambda_s/3$ from the free surface. 
An explosion is a purely compressional source,
and does not excite any $SH$ modes. Among the Lamb modes it excites preferentially the modes
that are``bulk" for both compressional and shear components as well as Rayleigh modes. A
source at a larger depth will not excite the Rayleigh modes since 
they have a penetration length of the order of the wavelength.
  
Figure~\ref{EnergyModes}$b$ applies 
for a double couple in the $xy$ plane at a depth $\lambda_s/3$ from the free surface.
Contrary to the isotropic explosion, the double couple in the $xy$ plane strongly excites the $SH$
modes. Since the source is close to the free surface Rayleigh modes are excited as well. The Lamb
modes which are ``bulk" for the shear component but only evanescent for the compressional component are
also excited.

Figures~\ref{EnergyModes}$c$, $d$ show the mode distribution for a double couple in the plane 
$xz$ for two different depths of the source,  $\lambda_s/3$ and $5\lambda_s$. When the source is
located close to the free surface the majority of the energy is distributed among the Rayleigh
modes.  Two Rayleigh modes are out of scale in Figure~\ref{EnergyModes}$c$ and carry half of the
released energy. Alone they carry
half of the total energy released.  On the other hand, when the source 
is situated deep in the plate the pattern becomes very
rich. One can see that the Rayleigh modes are not excited anymore.

\section{Diffusion Approximation}

Despite the many simplifications that have been carried out,
the final
transport equation (\ref{finalert}) is still difficult to solve numerically.
In future work, we intend to adapt our  Monte-Carlo simulations, 
developed to solve the 3D radiative transfer equation \cite{ludo1,ludo2},
to this  modified equation. In this section
we shall carry out a final  and rather familiar 
simplification, that   facilitates  a numerical solution.
 
 The diffusion approximation is typically valid at large lapse times,
 when currents start to become small. In that case, the specific intensity
 of mode $i$ can be written as,
 \begin{equation}
 L_{i{\mathbf k}}({\mathbf q},\Omega) = {1\over n_i} \left[ E_i ({\mathbf q},\Omega)
 + {2\over v_i^2} {\mathbf v}_i \cdot {\mathbf J}_i 
 ({\mathbf q},\Omega) + \cdots \right]\, , \label{da}
 \end{equation}
 with $n_i = k_i/v_i$ the density of mode $i$ in phase space introduced
 earlier. 
In real space ${\mathbf q}$ transforms into the 2D gradient $\nabla$. Inserting
the series (\ref{da}) into Eq.~(\ref{finalert}) leads to the relation
\begin{equation}
{\mathbf J}_i ({\mathbf r},t) = -\sum_j D_{ij}  \nabla E_j({\mathbf r},t)\, .
\label{fick}
\end{equation}
This relation is recognized as a generalized Fick's Law \cite{fick},
generalized, because it involves different individual modes at the
cost of one dimension. The {\it diffusion matrix} is given by,

\begin{eqnarray}
\left({\mathbf D}^{-1} \right)_{ij} = 2 \left( {\delta_{ij}\over v_i^2\tau_i}
- {\omega^2 \over n_j} \int {\mathrm d^2{\mathbf \hat{k}}_j\over 2\pi}\,  
W( i{\mathbf k}_i, i{\mathbf k}_j)
{{\mathbf v}_i \cdot {\mathbf v}_j \over v_i^2v_j^2} \right)\, , \label{diff}
\end{eqnarray}
It is easy to check the following relation,
\begin{equation}
{D_{ij} \over D_{ji} } = {n_i\over n_j}. \label{db}
\end{equation}
Combining Eqs.~(\ref{fick}) and (\ref{cont}) and transforming back 
to space-time yields the generalized 2D diffusion equation,
\begin{equation}
\partial_t E_i({\mathbf r},t) - \sum_j D_{ij}(\omega) \nabla^2 E_j({\mathbf r},t) 
=  S_i(\omega)   \delta(t) -\sum_j C_{ij}(\omega) 
E_j ({\mathbf r},t). \label{gendiff}
\end{equation}
This diffusion equation is an ordinary partial differential equation
that can be solved by conventional means. For an infinite plate no boundary 
conditions have to be specified: the boundary conditions
at the two free surfaces have been taken care of exactly.  For
this reason, the Q2D diffusion approximation is not expected to
break down near the boundaries, as was noticed by Turner and Weaver 
for the conventional diffusion approximation \cite{turnweav2}.   

Equation~(\ref{gendiff}) still captures the time-evolution
of the different elastic modes of the plate, 
and can thus be used to study polarization properties. 
Integrating equation~(\ref{gendiff}) over the horizontal coordinate ${\mathbf r}$ 
gives for the time evolution of the total modal energy
\begin{equation}
\partial_t E_i(t) 
=  S_i(\omega)   \delta(t) -\sum_j C_{ij}(\omega) 
E_j (t).
\label{intdiffeq} 
\end{equation}
In fact this equation  follows directly from Eq.~(\ref{cont}) without 
the need to apply the diffusion approximation.
Its formal solution is
$E_i(t)=\sum_j \left[\exp{(-{\mathbf C}t)} \right ] _{ij}
S_j(\omega)\theta(t>0)$. This can easily be evaluated using the
complete set of eigenmodes of ${\mathbf C}$.

Figure~\ref{EnergyTime}$a$ shows the time evolution of the energy among the different
modes for an isotropic explosion at a depth $\lambda_s/3$ from the free surface.
The initial modal energy distribution was shown in Figure \ref{EnergyModes}$a$. 
For the sake a clarity we only display the evolution of three sub-classes of modes
(Rayleigh, Lamb, SH) and not the whole distribution. 
Rayleigh modes are excited but not SH modes 
since the source is purely compressional. As time goes on, the mode occupation changes as a result of
the dynamics of the equipartition process and finally tend to the equipartitioned distribution
which does not depend on nature and location of the source.

Figure~\ref{EnergyTime}$b$ shows the time evolution of two ``observable''
energy ratios measured at the free surface:
the ratio of shear to compressional potential energy, $E_s/E_p$, 
and the one of the horizontal to vertical kinetic energy $H^2/V^2$. After a  few 
shear wave mean free times, the energy ratios stabilize to 
their predicted equipartition value $E_s/E_P=7.19$, $H^2/V^2=1.77$  \cite{hennino}. 
The ratios $E_s/E_P$ and $H^2/V^2$ increase monotonically which is due to
the compressional nature of the source.

Figures~\ref{EnergyTime}$c$, $d$ present the equipartition process 
for a double couple source deep in the plate ($5\lambda_s$ from the free surface).
 For such a source the Rayleigh modes are not excited 
while the other  Lamb modes and SH modes are strongly excited (see Figure~\ref{EnergyModes}$d$).
The initial ratio of shear to compressional energy at the free surface is higher than 
the one for the explosion source due to the 
shear nature of the source. However, in both cases the energy distributions
converge towards an equipartitioned distribution which is 
independent of the nature of the source and its location.
Note that, for an exploding source, the equipartition process takes much longer a time,
 typically $6\tau_s^\infty$. For the double-couple source in Figures~\ref{EnergyTime}$c$, $d$
 it is typically $\tau_s^\infty$.

It is not very difficult to show that in the equipartition regime,
the generalized diffusion equation (\ref{gendiff}) further simplifies
to a genuine 2D diffusion equation for the total energy density,
\begin{equation}
\partial_t E({\mathbf r},t) - D(\omega)\nabla^2 E({\mathbf r},t) =  S(\omega) 
  \delta({\mathbf r}) 
\delta(t)
\, , \label{ordd}
\end{equation}
with diffusion constant,
\begin{equation}
D(\omega) = {\sum_{ij} D_{ij}(\omega) n_j(\omega) \over \sum_j n_j(\omega) }\, ,\label{d3}
\end{equation}
and source,
\begin{equation}
S(\omega) = \sum_{i} n_i 
\int\mathrm {d^2 {\mathbf \hat{k}} \over  2\pi }\, \left|S_{i{\mathbf k}_i}(\omega)\right|^2
 \, .
\end{equation}
Equation~(\ref{d3}) is recognized as an equipartitioned sum of all diffusion matrix elements.
A similar result was obtained for the diffusion constant
in an infinite elastic medium, in terms of the individual
matrix elements for $P$ and $S$ waves \cite{weaver,turner,poan}.
Equation~(\ref{ordd}) has the simple solution,
\begin{equation}
E(\omega,{\mathbf r}, t) =   {  S(\omega)\over 4\pi D(\omega)t} 
\exp\left(-{{\mathbf r}^2\over
4 D(\omega)t }\right)\, , \label{drt}
\end{equation}
i.e.  the local energy basically varies
as
$E(\omega) \sim t^{-1} \times S(\omega)/D(\omega) $ at large times.

Table~\ref{cstdiff} shows the evolution of the ratio $D(\omega)/D^\infty(\omega)$ as a function of the
number of modes in the plate. $D^\infty(\omega)$ is the infinite medium elastic diffusion constant, obtained
by Weaver~\cite{weaver} and Ryzhik~\cite{ryzhik}, with the same amount of disorder, $i.e.$, 
as described by Eqs.~(\ref{correl}). 
The ratio changes form $0.72$ for $N=3$ modes to $0.85$ for a thick plate, $i.e.$ 
$H\approx l^*$. Our quasi-2D approximation starts to break down when the thickness of the plate
exceeds the mean free path.

\section{Coherent Backscattering near the Free Surface}\label{sectioncone}
 
Coherent backscattering of waves is an interference effect that
 survives multiple scattering. It refers to a coherent
  enhancement of intensity near the source \cite{poan}. The effect
  has recently been observed with acoustic \cite{coneacou} and
  elastic waves \cite{julien,r12}.

 We recently investigated coherent backscattering of waves in a more seismic
 context \cite{coneludo,coneik}. 
 Specific aspects such as symmetry of the source, near field, leakage
 and the exact measurement process have to be understood before
 any seismic experiment can be considered. Our analyses so far have been
 done either with scalar (acoustic) waves in a disordered 
 plate with leakage \cite{coneludo}
 or with fully elastic waves in an infinite medium \cite{coneik}. 
The last study established that the enhancement factor 
of coherent backscattering is highly dependent on both the nature of
the source and on the precise parameter that is being measured. More specifically,
a measurement of simply $\left<u_i(\omega)^2 \right>$ of waves released
by a ``double-couple" source will hardly give rise to a coherent enhancement,
so that observation is unlikely. On the other hand, the measurement
of $\left<\mathrm{div}\, {\mathbf u} (\omega)^2 \right>$ of waves released by
an explosion source
maps exactly onto the acoustic problem, which has the maximal 
enhancement factor of $2$.

Both approaches are unable  to model the coherent backscattering 
effect of wave propagation 
in the crust, whose elastic eigenmodes are not plane waves. In addition,
a measurement necessarily takes place at the Earth surface, whereas
the source (an earthquake or explosion) can be located at depth.
In this section we will investigate coherent backscattering using
our Quasi 2D transport model. Recently, De Rosny et al. \cite{julien}
and Weaver et al. \cite{r12} reported the studies of
coherent backscattering of elastic waves at frequencies around 1 MHz.

Our analysis will closely follow the one given in Ref.~\cite{coneik}.
Starting point is the calculation of the vertex
$L_{nn'mm'}({\mathbf k}, {\mathbf k}', {\mathbf q})$ defined in
Eq.~(\ref{si}) and  describing the ensemble-averaged, incoherent 
scattering of the modes $(i,{\mathbf k}+\frac12 {\mathbf q})$ and 
$(i',{\mathbf k}-\frac12 {\mathbf q})$ into   
$(j,{\mathbf k}'+\frac12 {\mathbf q})$ and $(j',{\mathbf k}'-\frac12 {\mathbf q})$. By the reciprocity
principle this object must be symmetrical with respect to left and right-hand
indices. The diffusion approximation, applied to
our Q2DA model yields for large lapse times,
\begin{equation}
 L_{ii'jj'}({\mathbf k}, {\mathbf k}', {\mathbf q}) = {\delta_{ii'} 
 \delta\left(\omega
 -\omega_{i{\mathbf k}} \right) \delta_{jj'} \delta\left(\omega
 -\omega_{j{\mathbf k}'} \right) \over -i\Omega + D{\mathbf q}^2 + \omega/Q}\, .
 \label{llbegin}\end{equation}
An inverse Fourier transform with respect to $\Omega$ provides the time-dependence
of the envelope of a wave packet with central frequency $\omega$
\footnote{Seismic measurements usually have access to
time-correlations $\left<\psi_j(t-\frac12 \tau) \psi_j^*(t+\frac12 \tau)\right>$
of different components of the wave function (\ref{psi}). The
smooth time-dependence of the envelope of the pulse at frequency $\omega$ 
can be obtained
by Fourier transforming this object with respect to $\tau$. Signal-to-noise
can usually be increased by averaging over a time window $\Delta t$ and 
using a finite bandwidth $\Delta \omega$ over which the signal is not 
expected to vary too much.}.   
Similarly, the spatial dependence is obtained by an inverse 
Fourier transform
over ${\mathbf q}$, ${\mathbf k}$ and ${\mathbf k}'$. The result  is,

\begin{equation}
L_{ii'jj'}(\omega, t,{\mathbf x}_1, {\mathbf x}_2 \rightarrow {\mathbf x}_3, {\mathbf x}_4)
= {\exp(-\omega t/Q) \over Dt} \, \delta_{ii'}\delta_{jj'} \, 
n_i  n_j  \, J_0\left(k_i x_{12} \right)J_0\left(k_j x_{34} \right)\, .
\label{LL} \end{equation}
The depth (i.e. $z$) dependence can by obtained by summing over the $N$
 eigenfunctions
${\mathbf \Psi}_i(z)$ at frequency $\omega$.
The coherent backscattering  is due to constructive interference of
time-reversed waves.  It can be constructed straightforwardly by interchanging
the  indices $(i'{\mathbf x}_2)$ and $(j'{\mathbf x}_4)$ \cite{coneik},
 \begin{equation}
C_{ii'jj'}(\omega, t,{\mathbf x}_1, {\mathbf x}_2 \rightarrow {\mathbf x}_3, {\mathbf x}_4)
= {\exp(-\omega t/Q) \over Dt} \, \delta_{ij'}\delta_{ji'} \, 
n_i n_j  \, J_0\left(k_i x_{14} \right)J_0\left(k_j x_{32} \right)\, .
\label{CC}\end{equation}
Both $L$ and $C$ contribute to 
$\left< G(i,{\mathbf x}_1\rightarrow j,{\mathbf x}_2)
G^*(i',{\mathbf x}_3\rightarrow j',{\mathbf x}_4) \right>$, but 
$C$ survives only close to the source, as we shall see.
To calculate actual enhancement factors, we must specify   source
and   detector. In Eq.~(\ref{source}) the source was already 
expressed in the eigenmodes
$(j,{\mathbf k})$. Different sources will now be considered.

\subsection{Monopolar source at depth}

We consider the source ${\mathbf f} ({\mathbf r}) \sim {\mathbf f}_0(\omega) 
\delta^{(3)}
({\mathbf r} - {\mathbf r}_0)$, which represents a highly directional force field
at position ${\mathbf r}_0$,
small compared to the wavelength.  Equation~(\ref{source}) gives 
$S_{j{\mathbf k}}(\omega) \sim 
\omega {\mathbf f}_0(\omega) \cdot
{\mathbf u}_{j{\mathbf k}_j}(z_0)$ with $z_0$ the depth of the source.
To simplify the analysis we will assume that the force is directed into
the $z$-direction. This configuration was also studied by
De Rosny et al. \cite{julien,thesejulien} using a thin chaotic 2D silicon cavity,
with only 3 excited Lamb waves. In addition, their detection method of 
 heterodyne laser interferometry  is only sensitive to 
the normal displacement $u_z(z=0)$. In seismology,
the force field above may be a simple model for a volcano eruption.

Let  ${\mathbf x}$ be the horizontal distance between source and detector.
The measured ``incoherent" background is found from Eq.~(\ref{LL}),
\begin{equation}
L(x,t) \sim
 {\exp(-\omega t/Q) \over Dt} \, f_0(\omega)^2 \sum_i n_i  
   \left|u_{i,z}(0) \right|^2  \sum_j n_j    
 \left| u_{j,z}(z_0)\right|^2 \, , 
\end{equation}
which is independent of ${\mathbf x}$, but still depends on
the depth $z_0$ of the source.
The "coherent" contribution follows from Eq.~(\ref{CC})
\begin{equation}
C(x,t) \sim
 {\exp(-\omega t/Q) \over Dt} \, f_0(\omega)^2 
 \left| \sum_i n_i \,   
   u_{i,z}(0)  
 \, u_{i,z}(z_0)^*  J_0(k_i x) \right|^2  \, . 
\end{equation}
As was already mentioned in previous work, 
the ratio $(L+C)/L$, the so-called ``enhancement factor", is
independent of time at large lapse times \cite{coneludo}.
An application of Cauchy's inequality shows that
$(L+C)/L \leq 2$, with equality {\it only} if $x=0$ and if
$ u_{i,z}(0)  =   
 u_{i,z}(z_0)$ for all modes~$i$. This can only be true
 if $z_0 = 0$ i.e.
 the source must be near the surface. In practice, 
 to produce any measurable enhancement factor, the source 
 must be at a depth less than the typical wavelength, as shown in 
 Figures~\ref{coneMP}$a$ and \ref{coneMP}$b$.
 A source with a force direction different from normal would 
 have a lower enhancement as well. 
 Note that the enhancement is symmetric in azimuth 
 around the source.

\subsection{Isotropic Explosion}
 
 An isotropic explosion at depth $z_0$ is described by the force field
 ${\mathbf f}({\mathbf r}, \omega) = B(\omega) \nabla   
 \delta({\mathbf r}-{\mathbf r}_0)
 $ \cite{seis2}. It can easily be shown that $S_{i{\mathbf k}} (\omega) =
 -B(\omega) \omega \, \mathrm{div}\, {\mathbf u}_{i{\mathbf k}}(\omega) $. 
 For a fixed frequency this
 depends on the mode label $i$ but, very conveniently, 
 not on the direction ${\mathbf \hat{k}}$ of
 horizontal propagation.   
 
 Let us first suppose that we measure the normal component of the 
 displacement vector at the surface. Incoherent background and coherent enhancement 
 are given by,
 \begin{eqnarray}
L(x,t) &\sim&
 {\exp(-\omega t/Q) \over Dt} \,    B(\omega)^2 \sum_i n_i  
 \, \left|u_{i,z}(0)\right| ^2  \, \sum_j n_j  
  \left| \mathrm{div}\, {\mathbf u}_{j}(z_0) \right| ^2      \, , \nonumber \\
C(x,t) &\sim&
 {\exp(-\omega t/Q) \over Dt} \,   B(\omega)^2 \left| 
 \sum_i n_i \,  
  u_{i,z}(0) \mathrm{div}\, {\mathbf u}_{i}(z_0)^* 
   J_0(k_i x) \right|^2  \, . 
\end{eqnarray}
The resulting enhancement factor $(L+C)/L$ is plotted in dashed lines in
Figure~\ref{coneEXP}$a$ 
as a function of the horizontal distance, and in Figure~\ref{coneEXP}$b$ for
a measurement on top of an explosion source as a function of the depth $z_0$. Note that
the enhancement never reaches its maximum value $2$, not even when $z_0=0$.
In an infinite medium, a measurement of any component of the displacement
vector of waves released by an explosion source would have had
no enhancement at all near the source \cite{coneik}. Here, the finite
enhancement is due to the nearness of a free surface.

 The enhancement factor can be restored by a measurement of the  
 dilatation ($\mathrm{div}\, {\mathbf u} $) in  which case,
 \begin{eqnarray}
L(x,t) &\sim&
 {\exp(-\omega t/Q) \over Dt} \,    B(\omega)^2 \sum_i n_i  
 \left| \mathrm{div}\, {\mathbf u}_{i}(0) \right|^2 
 \, \sum_j n_j  
  \left|\mathrm{div}\, {\mathbf u}_{j}(z_0)\right|^2      \, , \nonumber \\
C(x,t) &\sim& B(\omega)^2
 {\exp(-\omega t/Q) \over Dt} \,   \left| 
 \sum_i n_i  
 \mathrm{div}\, {\mathbf u}_{i}(0)    \mathrm{div}\, {\mathbf u}_{i}(z_0)^* 
   J_0(k_i x) \right|^2  \, .\end{eqnarray}
A measurement of the dilatation 
restores the symmetry between detector
and source, and reveals the maximum enhancement factor
$2$ when the detector is located close to the source as shown in solid lines in
Figures \ref{coneEXP}$a$, $b$.

\subsection{Dipolar Source}

We next consider a single couple at the surface with normal displacement
vector, and axis along the $x$-axis. This source can be 
represented by the dipole ${\mathbf f}({\mathbf r}, t) \sim
d(\omega)\, {\mathbf \hat{z}}   \partial_x \delta^{(3)} ({\mathbf r}-{\mathbf r}_0)$.  
Such a source can be  generated with laser interferometry 
on an elastic plate, and the resulting
coherent backscattering effect was
recently studied experimentally by De Rosny et al. \cite{thesejulien}.

The spatial derivative in the source finds its way in the Bessel functions,
in the same way as was done in earlier work  for the infinite system \cite{coneik}.
We derive, again for a measurement of the displacement vector
in the direction normal to the surface,
\begin{eqnarray}
L({\mathbf x},t) &\sim&
 {\exp(-\omega t/Q) \over Dt} \,\frac12  d(\omega)^2 \sum_i n_i  
 \, \left|u_{i,z}(0)\right| ^2  \, \sum_j n_j  
  \left|u_{j,z}(z_0)\right|^2  k_j^2    \, , \nonumber \\
C({\mathbf x},t) &\sim&
 {\exp(-\omega t/Q) \over Dt} \, \cos^2\phi\, d(\omega)^2  \left| 
 \sum_i n_i k_i 
  u_{i,z}(0) 
 u_{i,z}(z_0)^* 
   J_1(k_i x) \right|^2  \, . 
\end{eqnarray}
Two things can be noted. First, $C$ vanishes anywhere above the source,
(${\mathbf x}=0$). The enhancement is destroyed because the dipolar nature of the
source is in some sense ``orthogonal" to the detection of  the displacement
vector. Maximum enhancement actually occurs a fraction of a 
wavelength away from the source as shown in dashed line in Figure~\ref{coneDIPOLE}$a$.
Secondly, the coherent enhancement around the source has
a $\cos^2 \phi$ structure, with $\phi$ the azimuthal angle between
the dipole-axis of the source and the direction
 of detection. This ``double-well" structure
was observed by De Rosny, Tourin and Fink \cite{thesejulien}.

The coherent enhancement factor can be restored by a  modification
of the measurement. Suppose we measure the parameter
$ \partial_x u_z({\mathbf r},t)$. 
This measurement has the same symmetry as the dipolar source.
We find for background and coherent enhancement,
\begin{eqnarray}
L({\mathbf x},t) &\sim&
 {\exp(-\omega t/Q) \over Dt} \,\frac14  d(\omega)^2 \, \sum_i n_i  k_i^2
  \left|u_{i,z}(0)\right|^2   \sum_j n_j k_j^2 
 \left|u_{j,z}(0)\right|^2      \, , \nonumber \\
C({\mathbf x},t) &\sim&
 {\exp(-\omega t/Q) \over Dt} \times \nonumber \\ 
 &\,& d(\omega)^2\left| \sum_i n_i k_i^2  
 u_{i,z}(0)  
  u_{i,z}(z_0)^* 
 \left[ {J_1(k_ix) \over k_ix}  
 - J_2(k_ix) \cos^2\phi \,
 \right]
  \right|^2  \, . 
\end{eqnarray}
For $x=0$ and $z_0 =  0$ we infer that $L=C$, 
i.e the maximal enhancement can now be reached. The plot of the restored
enhancement factor as the function of the horizontal distance and as the 
function of the source depth are shown in solid line in the Figures \ref{coneDIPOLE}$a$, $b$.
 Note that the line profile is
still not cylindrically symmetric, but depends on $\phi$.

\subsection{Double-couple source at depth}

Seismic sources have successfully been modeled as two compensating
couples (dipoles) \cite{akibook}. 
To facilitate observation of coherent backscattering
with seismic waves we will here obtain the enhancement expected
for such a source close to a free surface. In view of the complexity of the problem, we
will restrict ourselves to a seismic plane that is oriented parallel to the
free surface where detection takes place. The depth of this plane is located at
$z_0$.

The force field of a double-couple source is described by a symmetric,
 off-diagonal
seismic tensor. We assume that the two dipoles are   orthogonal
 and along the axes $x$ and $y$. The force field
is then given by,
\begin{equation}
{\mathbf f} ({\mathbf r}, \omega) = M(\omega) \left( {\mathbf \hat{x}}  \partial_y
+ {\mathbf \hat{y}} \partial_x  \right) \delta ({\mathbf r} - {\mathbf r}_0)
\, .\label{double}
\end{equation} 
with ${\mathbf r}_0 = (0,0,z_0)$. We can easily check that
the mode representation of the source (\ref{source}) is 
$S_{i,{\mathbf k}} = \omega M(\omega) \left[ k_x u_{i,y}(z_0) +k_y u_{i,x}(z_0)
 \right) $
 We will assume that
the measured parameter is  $    \partial_{y'} u_{x'}
+   \partial_{x'}  u_{y'}$, i.e. a certain horizontal component
of the stress tensor; $(x',y')$ are the  
coordinates in a frame
that has been rotated over an angle $\beta$ with $(x,y)$ 
(see Figure~\ref{angle}). The displacement vector of a mode
$(i{\mathbf k})$ can be expressed as,
\begin{equation}
{\mathbf u}_{i{\mathbf k}}(z) = \left\{ u_{i,z}(z) \, {\mathbf \hat{z}}
+ u_{i,\parallel}(z) \left[ \cos\alpha_i \, {\mathbf \hat{k}}
+ \sin \alpha_i   \, {\mathbf \hat{z}}\times {\mathbf \hat{k}} \right]
\right\} \exp(i{\mathbf k\cdot x})\, ,
\end{equation}
which introduces a new angle $\alpha_i$ independent  
on the direction  ${\mathbf k}$ of propagation and on depth. 
Lamb waves have $\alpha_i=0$ whereas SH-waves have $\alpha_i=\frac\pi 2$.
We define $\phi$ as the angle between ${\mathbf k}$ and the $x$-axis, i.e.
${\mathbf \hat{k}} = \cos\phi {\mathbf \hat{x}}+ \sin \phi {\mathbf \hat{y}}$.
Finally, the angle $\mu$ orientates the direction
of measurement ${\mathbf x}$ in the horizontal plane 
with respect to the source.
(see Figure~\ref{angle}). 

The incoherent background is calculated from,
\begin{eqnarray}
L({\mathbf x},t) \sim 
{\exp(-\omega t/Q) \over Dt} \, M(\omega)^2
\sum_i \int \mathrm d^3{\mathbf k}\, 
\left[ \partial_{x'}u_{i{\mathbf k}, y'}(0) 
+ \partial_{y'}u_{i{\mathbf k}, x'}(0) \right]^2\, 
\mathrm{Im}\, G_{i{\mathbf k}}(\omega)  \nonumber \\
 \times \sum_j \int \mathrm d^3 {\mathbf k}' \left[ \partial_{x}u_{j{\mathbf k}', y}({\mathbf r}_0)
+ \partial_{y}u_{j{\mathbf k}', x}({\mathbf r}_0)  
\right]^2\, \mathrm{Im}\, G_{j{\mathbf k}'}(\omega) 
\end{eqnarray}
whereas the coherent enhancement follows from,
\begin{eqnarray}
C({\mathbf x},t) &=& 
{\exp(-\omega t/Q) \over Dt} \, M(\omega)^2  \times \nonumber \\ 
&&\left|
\sum_i \int \mathrm d^3 {\mathbf k} \left[\partial_{x'}u_{i{\mathbf k}, y'}
({\mathbf 0}) 
+ \partial_{y'}u_{i{\mathbf k}, x'}(0) \right]
\left[\partial_{x}u_{i{\mathbf k}, y}
({\mathbf r}_0) 
+ \partial_{y}u_{i{\mathbf k}, x}({\mathbf r}_0) \right] 
\mathrm{Im}G_{i{\mathbf k}}(\omega) 
\right|^2.
\end{eqnarray}
These ${\mathbf k}$-integrals
 can be evaluated straightforwardly and we simply quote the final result,
 \begin{eqnarray}
L({\mathbf x},t) &\sim&
 {\exp(-\omega t/Q) \over Dt} \,\frac14  M(\omega)^2 \sum_i n_i k_i^2  
 \left| u_{i,\parallel}(0) \right|^2  \sum_j n_j k_j^2 
 \left| u_{j,\parallel}(z_0) \right|^2 \, .\end{eqnarray}
 Here, $u_\parallel$ denotes the  complex amplitude of the horizontal component
 of the displacement vector. The coherent part is,
 \begin{eqnarray}
C({\mathbf x},t)  &\sim& 
 {\exp(-\omega t/Q) \over Dt} \,\frac14 
    M(\omega)^2 \times \nonumber\\
  && \left| \sum_i n_i k_i^2  
 u_{i,\parallel}(0)  
  u_{i\parallel}(z_0)^*   
   \left[ \cos \beta J_0(k_ix) 
  - \cos q_i J_4(k_ix)\right]  \right|^2.\end{eqnarray}
with $q_i = 4\mu +3\beta$ for Lamb waves and $q_i = 4\mu +3\beta+\pi$ for $SH$ waves.
Since $J_4$ term is very small,
the line profile is almost isotropic around $x=0$, independent on
$\mu$ , and maximal for $\beta=0$. The enhancement factor $(L+C)/L$ is plotted 
in Figure~\ref{coneDOUBLECOUPLE}$a$ as a function of the horizontal distance for different
source depth and in Figure~\ref{coneDOUBLECOUPLE}$b$ as a function of the source depth $z_0$ 
for a measurement on the top of the source.
It is interesting to notice the relatively large second maximum of the coherent backscattering
for a source at $x=0.7\lambda_s$ away from the detector ($x\approx1.2km$).

\section{Conclusions and Outlook}\label{conclusions}

In this paper we have investigated multiple scattering of elastic waves for a model that is adapted
to the needs of seismology of the Earth's crust: a two-dimensional solid plate with a thickness that
is less than the mean free path of the waves. Contrary to other approaches, this model facilitates an
exact treatment of the boundary condition at both sides of the plate, $i.e.$ on the level of the
elastic waves equation, and allows for fluctuations in the elastic constants that are depth-dependent.
At the same time, we can describe the horizontal transport of waves, as well as the inter-mode
mixing, by a generalized radiative transfer equation, that can be solved with conventional methods.
Using this equation, we have investigated different aspects, such as surface detection, mode
extinction times, equipartition, polarized sources at different depths in the plate and coherent
backscattering. We believe that this study is an important step in the modelisation of seismic waves
in the Earth's crust, but may also find applications in laboratory experiments.
In future studies we will try to solve our equation numerically using Monte-Carlo methods. 

One final
limitation of the present model has to be looked at in more detail. In this paper we have assumed a
solid plate bounded by two ideal free surfaces without any leaks. The Earth's crust is much better
described by one top free surface and a solid-solid interface at 30km in depth (the so-called Moho)
overlying a high-velocity mantle. In  previous studies we already suggested that the resulting
energy leak into the mantle may be the origin of seismic coda~\cite{ludo1,ludo2}. A future study
should establish the relation between seismic coda and the individual quality factors of the modes.

\section*{Acknowledgment}
We are indebted to Michel Campillo, Ludovic Margerin, Renaud Hennino and Celine Lacombe for many
constructive discussions. This work was supported by the Groupement De Recherche ``PRIMA" of the
French CNRS.

\newpage

\begin{table}
\begin{tabular}{|c|cccccccc|}
$N(\omega)$& 3& 5&13& 23& 43& 65& 85& 106\\\hline
$\frac{D(\omega)}{D^\infty(\omega)}$ &0.72&0.56&0.72& 0.77& 0.82& 0.84&0.85&0.85\\
\end{tabular}\vspace{0.5cm}
\caption{Ratio $D(\omega)/D^\infty(\omega)$ as a function of the number of modes $N(\omega)$ in the plate.
$D^\infty(\omega)$ is the frequency-dependent diffusion constant for a 3D infinite medium,
$D(\omega)$ is the frequency-dependent diffusion coefficient for our quasi-2D model
with $N(\omega)$ modes, with the same kind of disorder in $\lambda$ and~$\mu$.}\label{cstdiff}
\end{table}

\newpage

\begin{figure}
\caption{Schematic plot of the dispersion law of the 
elastic Rayleigh-Lamb eigenmodes in a layer
bounded by two free surfaces. Bold lines indicate symmetric branches,
straight lines indicate anti-symmetric modes. Only modes of different symmetry
are allowed to cross. The two dashed lines indicate the
pure shear or pure compressional excitations. The surface Rayleigh waves
propagate somewhat slower than pure $S$ waves.}\label{modes}
\end{figure}

\begin{figure}
\caption{Extinction times for the different modes, calculated from Eq.~(\ref{tau}), normalized by the 
mean free time of $S$ waves in an infinite medium. The disorder is chosen to be uniform in the 
whole plate and the spatial correlation between the Lam\'e coefficients is chosen equal. 
The plate thickness is $H=20.2\lambda_s$, which has a number 
of $N=106$ modes.}\label{extime}
\end{figure}

\begin{figure}
\caption{
Eigenvalues of the matrix ${\mathbf C}$ defined in Eq.~(\ref{mcon}) in the case of a plate thickness 
$H=20.2\lambda_s$ having a number of $N=106$ modes normalized to the mean free time of $S$-waves in
an infinite medium. The disorder is uniform in the whole plate and the spatial correlation 
between all Lam\'e coefficients is equal.}\label{eigenC}
\end{figure}

\begin{figure}
\caption{Energy distribution among the different modes for different sources.
The total energy release is normalized to 1
and the plate thickness is $H=20.2\lambda_s$ with $N=106$ modes.
A)~Isotropic explosion source at a depth $\lambda_s/3$ from the free surface.
B)~Double couple source in the $xy$ plan at a depth $\lambda_s/3$ from the free surface.
C)~Double couple source in the $xz$ plan at a depth $\lambda_s/3$ from the free surface. 
The two Rayleigh modes are out of scale and carry half of the released energy.
D)~Double couple source in the $xz$ plan at a depth $5\lambda_s$ from the free surface.
}\label{EnergyModes}
\end{figure}

\begin{figure}
\caption{Prediction of the diffusion equation (\ref{intdiffeq}) for the time evolution of 
the energy for different sources. The time scale has been normalized to the mean free time of the
S waves in an infinite medium. The plate thickness is $H=20.2\lambda_s$, with $N=106$ modes.
A)~and C) are predictions for the evolution of the energy
for different modes, $SH$ modes $(E_{SH})$,
Lamb modes without the Rayleigh modes $(E_{(L-R)})$, and Rayleigh
modes alone $(E_{R})$ for  respectively an isotropic explosion at a depth $\lambda_s/3$
from the free surface and a double couple source  
in the $xz$ plan at depth $5\lambda_s$ from the free surface.
B)~and D) are predictions for the potential energy ratio
 $E_s/E_p$ and the ratio
$H^2/V^2$ of the kinetic energies for the elastic displacement in different
directions.}\label{EnergyTime}
\end{figure}

\begin{figure}
\caption{Plot of the backscattering cone and enhancement factor
for a monopolar source along the $z$  axis. The normal component of the displacement 
field $u_{z}(0)$ is measured at the free surface and the plate thickness is $H=20.2\lambda_s$
 which has $N=106$ modes.
A)~Plot of the backscattering cone for different depths $z_0$. B)~Plot of the 
enhancement factor at $x=0$ as a function of the source depth 
$z_0$.}\label{coneMP}
\end{figure}

\begin{figure}
\caption{Plot of the backscattering cone and enhancement factor
for a isotropic explosion source. Both the divergence
(solid line) and normal component of the field (dashed line) are measured.
The plate thickness is $H=20.2\lambda_s$ which has $N=106$ modes.
A)~Plot of the backscattering cone. B)~Plot of the enhancement factor at $x=0$ 
as a function of the source depth $z_0$.}\label{coneEXP}
\end{figure}

\begin{figure}
\caption{Plot of the backscattering cone and enhancement factor
for a dipolar source. Both $\partial_x u_z({\mathbf r},t)$
(solid line) and normal component $u_z({\mathbf r},t)$ of the field (dashed line)
are measured. The plate thickness is $H=20.2\lambda_s$ which has $N=106$ modes.
A)~Plot of the backscattering cone. B)~Plot of the enhancement factor at $x=0$ 
as a function of the source depth $z_0$.}\label{coneDIPOLE}
\end{figure}

\begin{figure}
\caption{All angles involved in the measurement of the backscattering
cone of a dislocation source at depth $z_0$. See text for discussion,
$\alpha_i=0$ for Lamb waves and $\alpha_i=\pi/2$ for $SH$ waves.}\label{angle}
\end{figure}

\begin{figure}
\caption{Plot of the backscattering cone and enhancement factor
for a double-couple source with both its axes along the free surface. 
The orientation of the detection is such that $\beta=0$ and $\mu=0$.
The plate thickness is $H=20.2\lambda_s$ which has $N=106$ modes.
A)~Plot of the backscattering cone for different source depths $z_0=0$, $z_0=\lambda_s/3$
and $z_0=\lambda_s/2$. B)~Plot of the enhancement factor at $x=0$ 
as a function of the source depth $z_0$.}\label{coneDOUBLECOUPLE}
\end{figure}

\newpage

\begin{figure}
{\psfig{figure=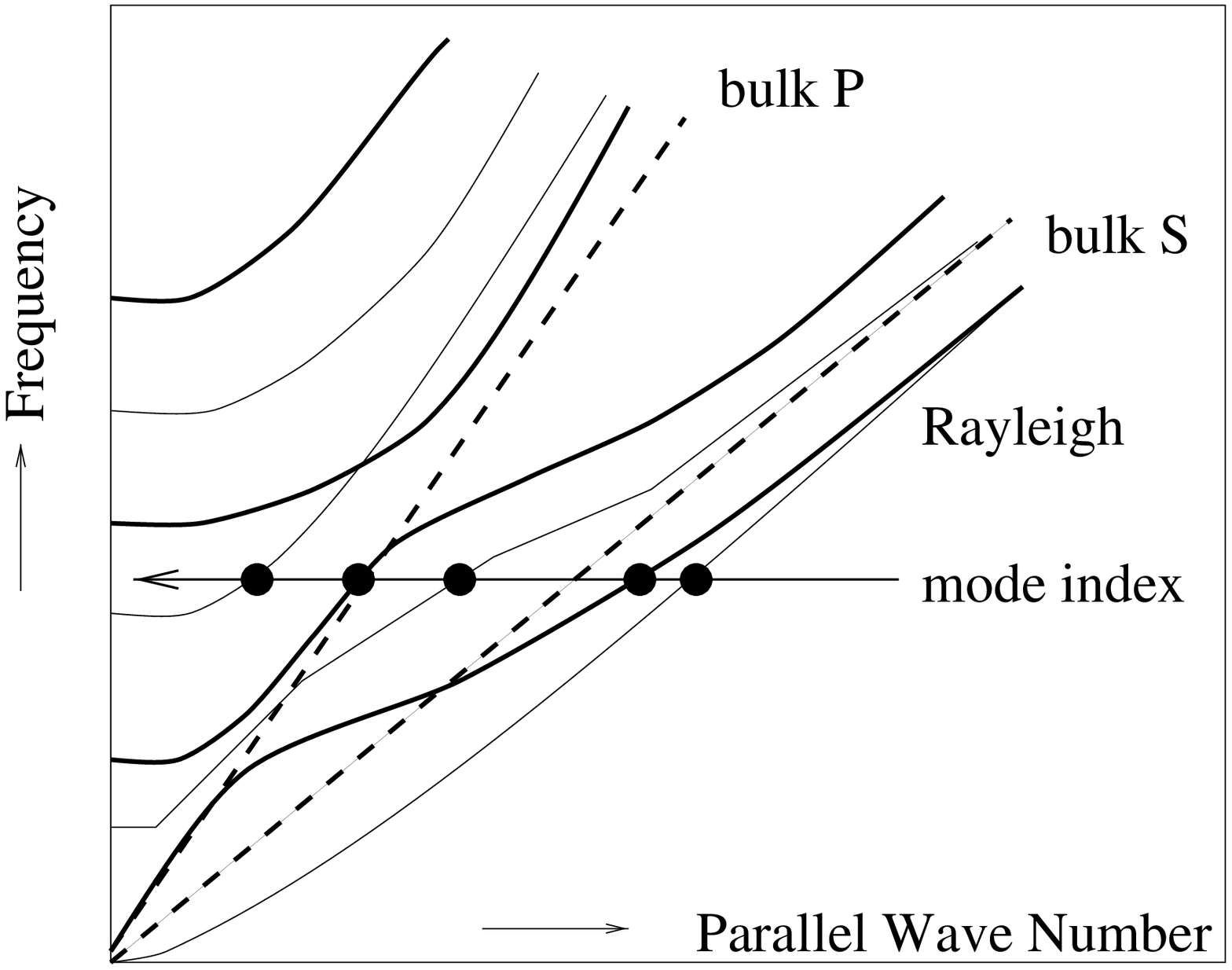,angle=-90,height=7.0cm,width=8.5cm}}
\end{figure}

\newpage

\begin{figure}
{\psfig{figure=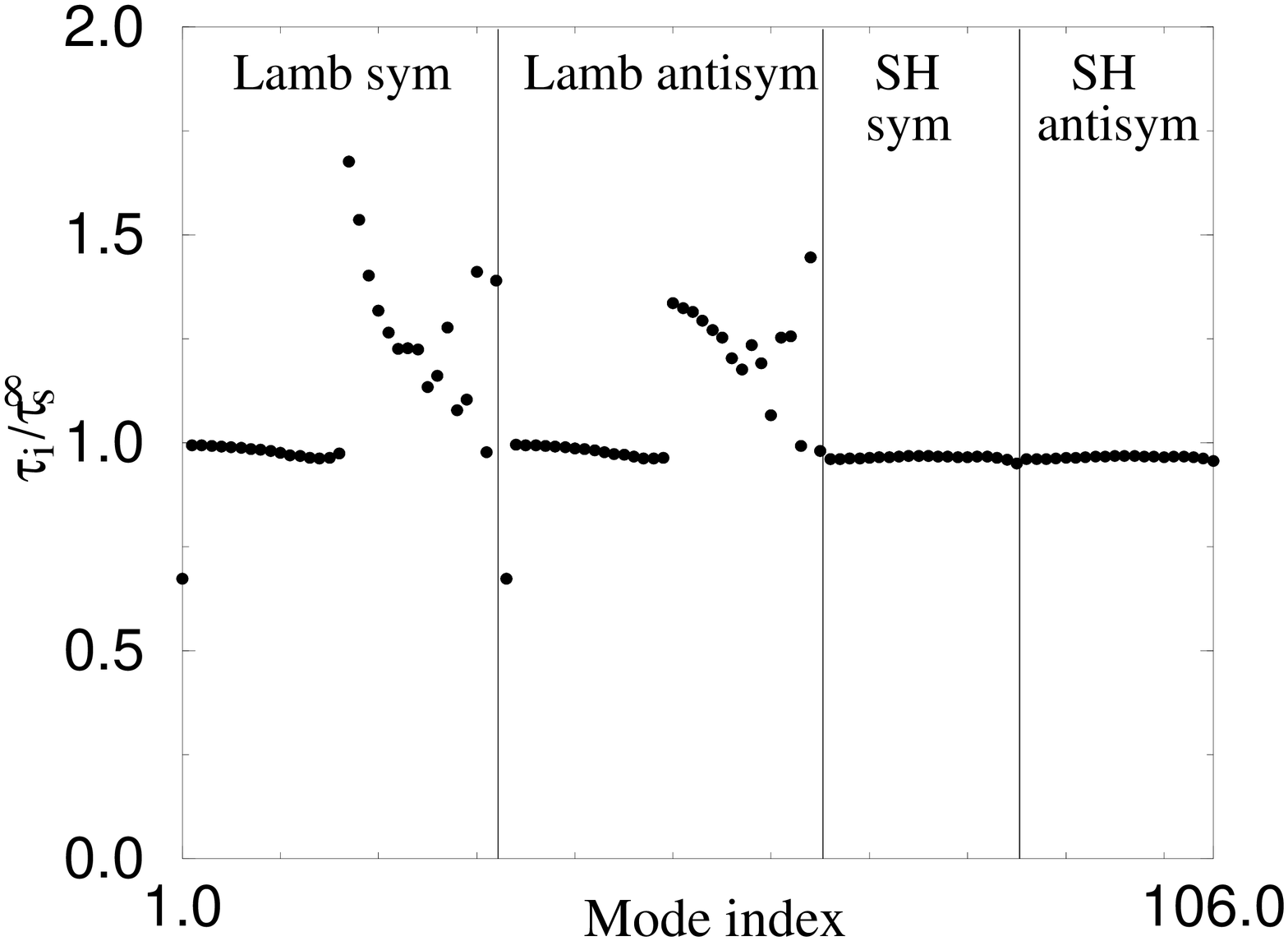,angle=-90,height=7cm,width=8.5cm}}
\end{figure}

\newpage

\begin{figure}
{\psfig{figure=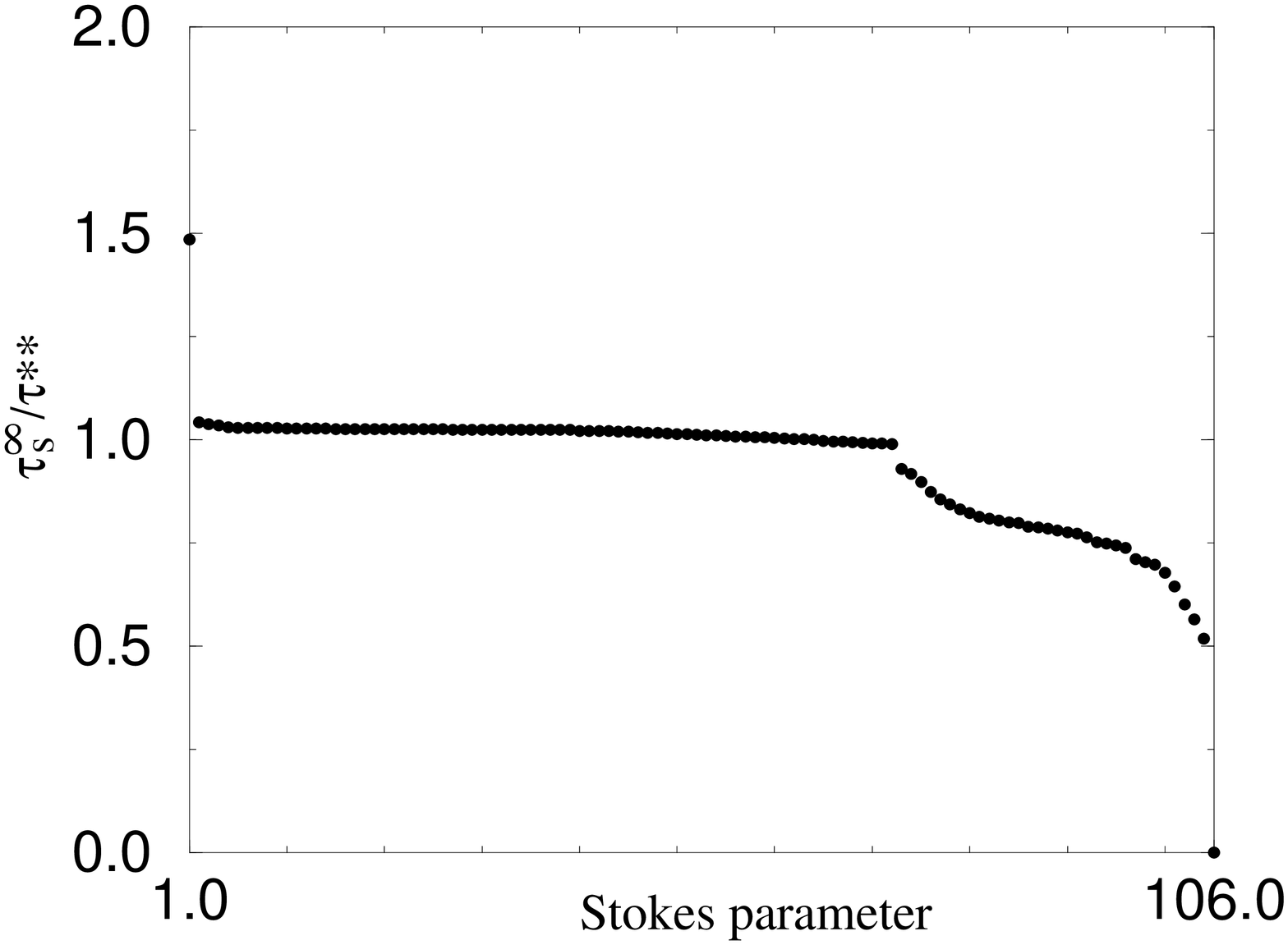,angle=-90,height=7cm,width=8.5cm}}
\end{figure}

\newpage

\begin{figure}
\noindent
\begin{minipage}[t][5cm][s]{8cm}
{\psfig{figure=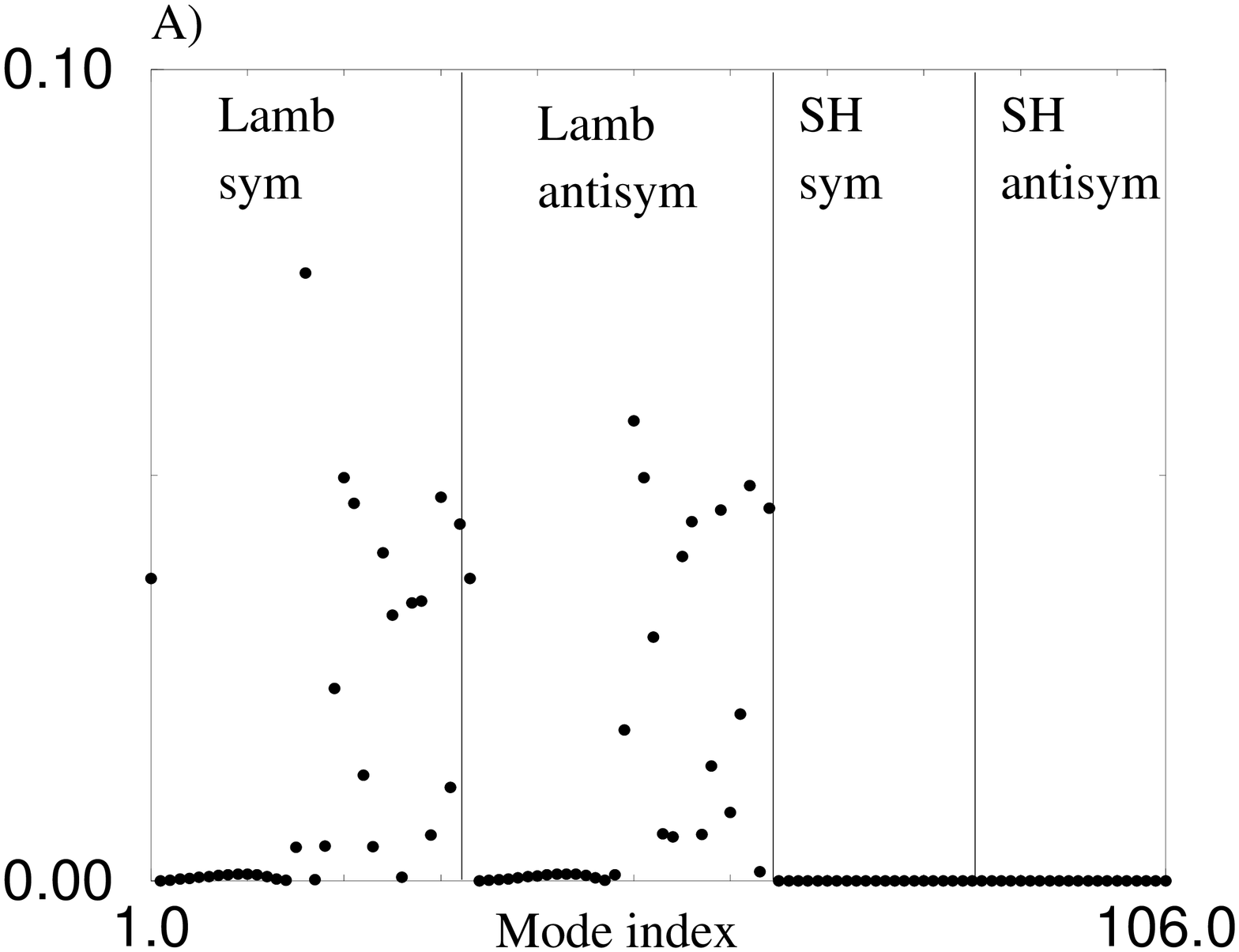,angle=-90,height=5cm,width=8cm}}
\end{minipage}\hfill
\begin{minipage}[t][5cm][s]{8cm}
{\psfig{figure=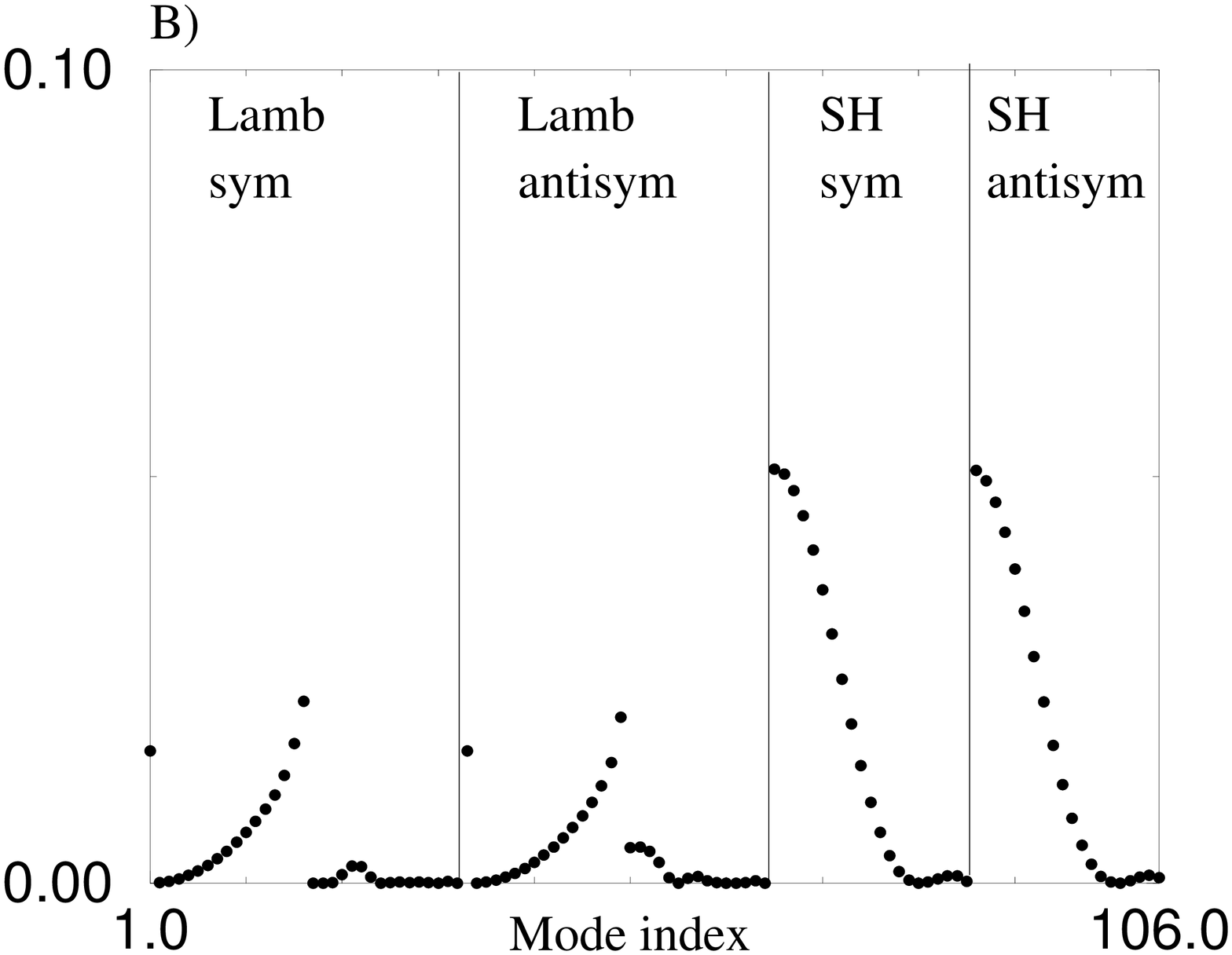,angle=-90,height=5cm,width=8cm}}
\end{minipage}
\begin{minipage}[t][5cm][s]{8cm}
{\psfig{figure=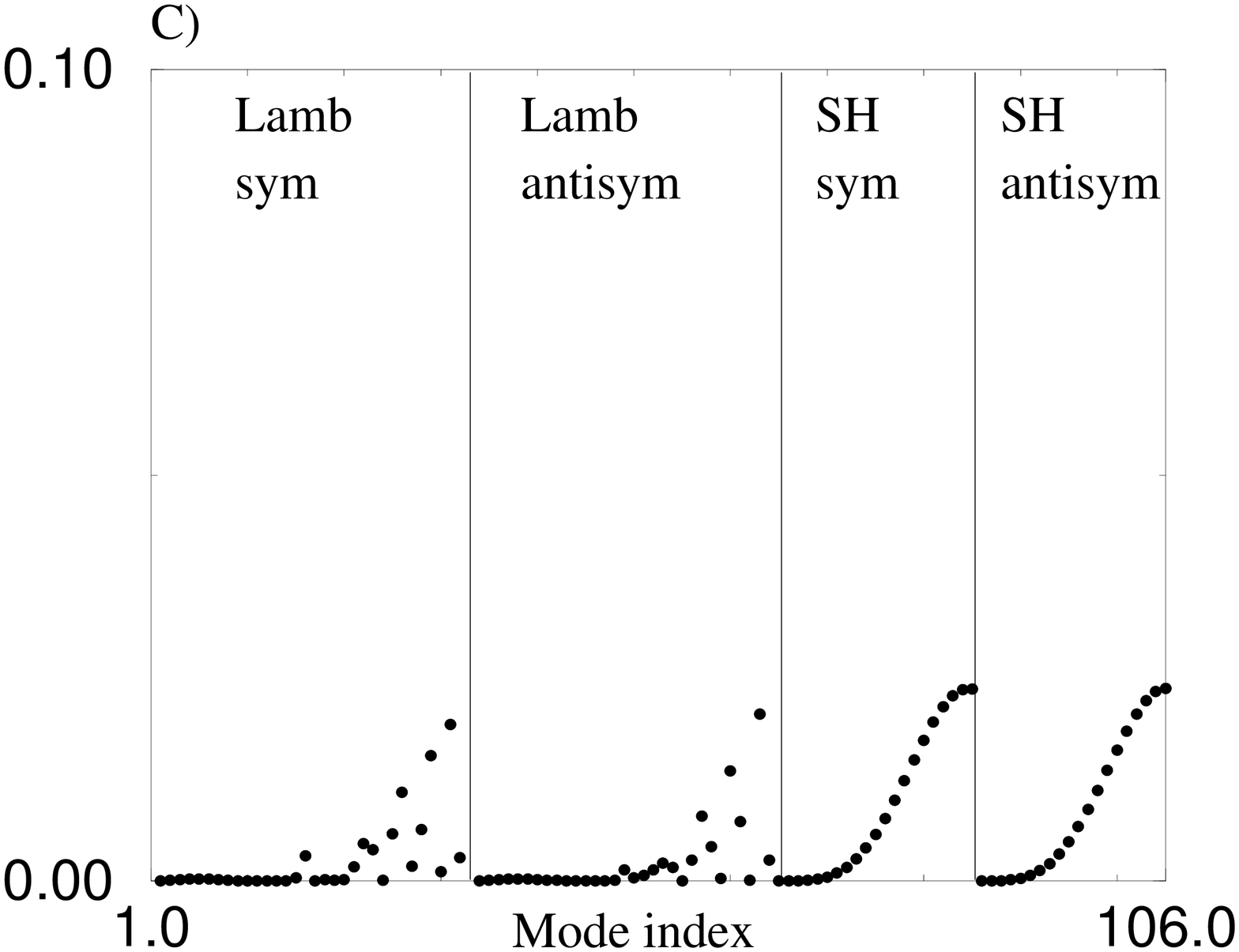,angle=-90,height=5cm,width=8cm}}
\end{minipage}\hfill
\begin{minipage}[t][5cm][s]{8cm}
{\psfig{figure=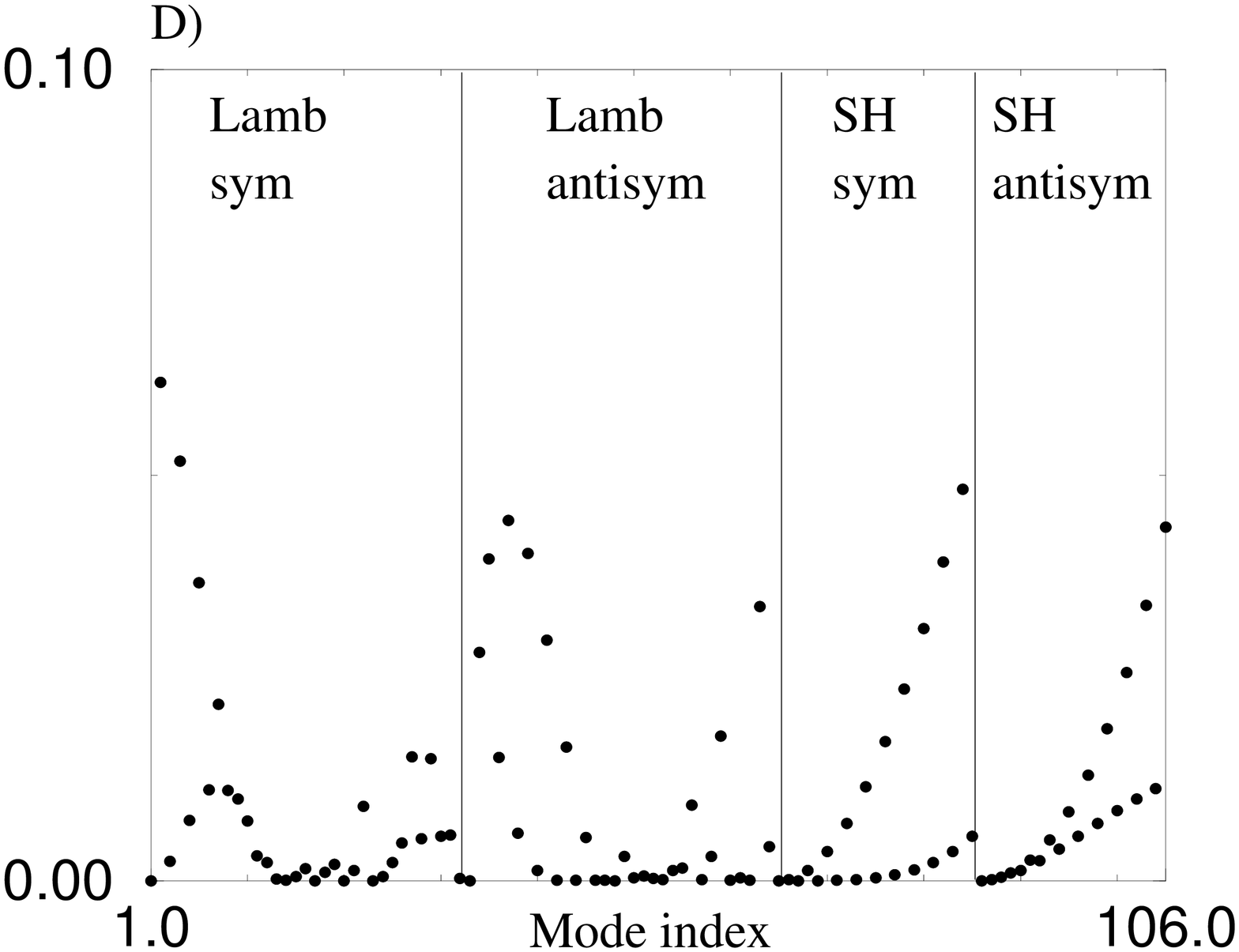,angle=-90,height=5cm,width=8cm}}
\end{minipage}
\end{figure}

\newpage

\begin{figure}
\noindent
\begin{minipage}[t][5cm][s]{8cm}
{\psfig{figure=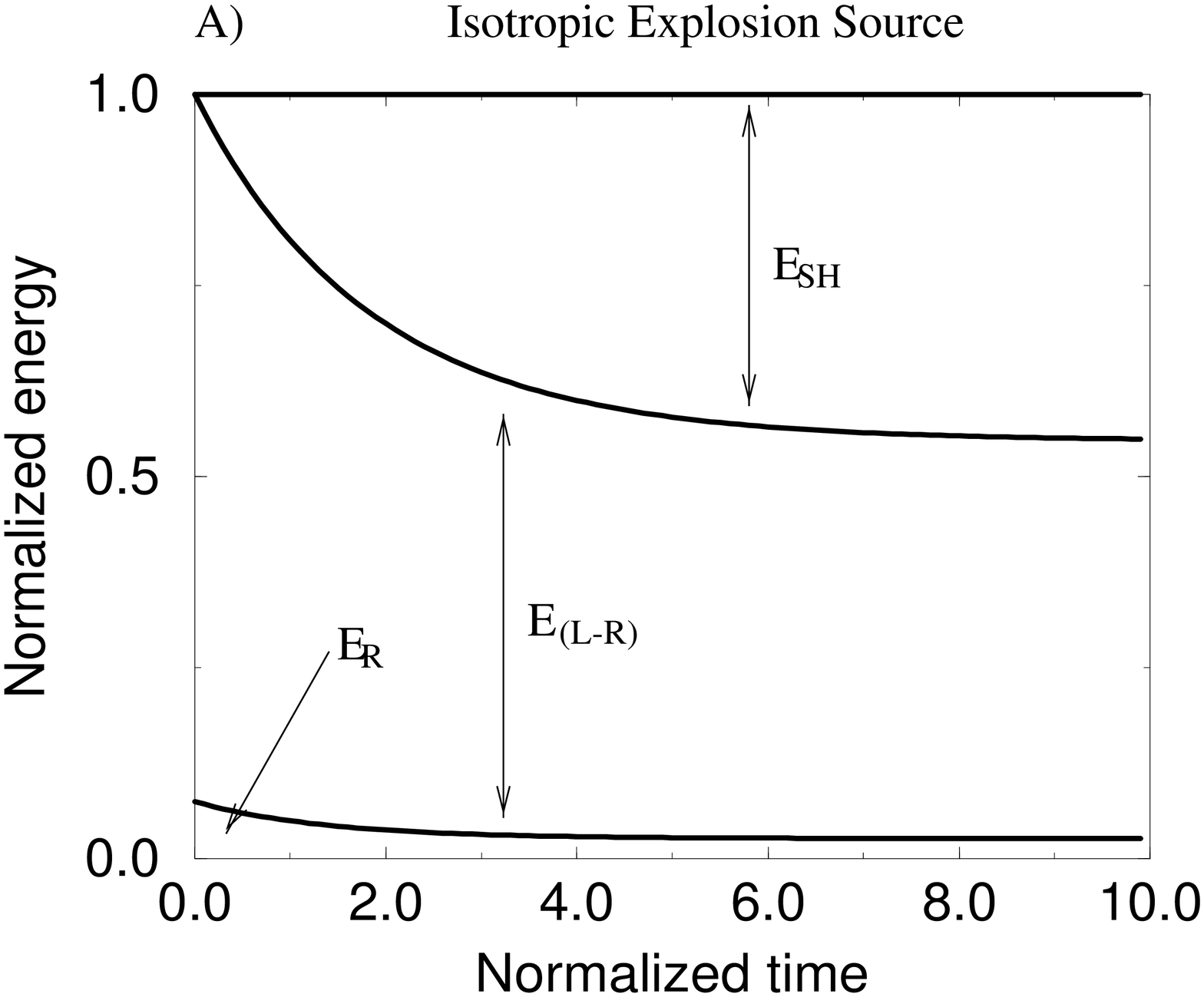,angle=-90,height=5cm,width=8cm}}
\end{minipage}\hfill
\begin{minipage}[t][5cm][s]{8cm}
{\psfig{figure=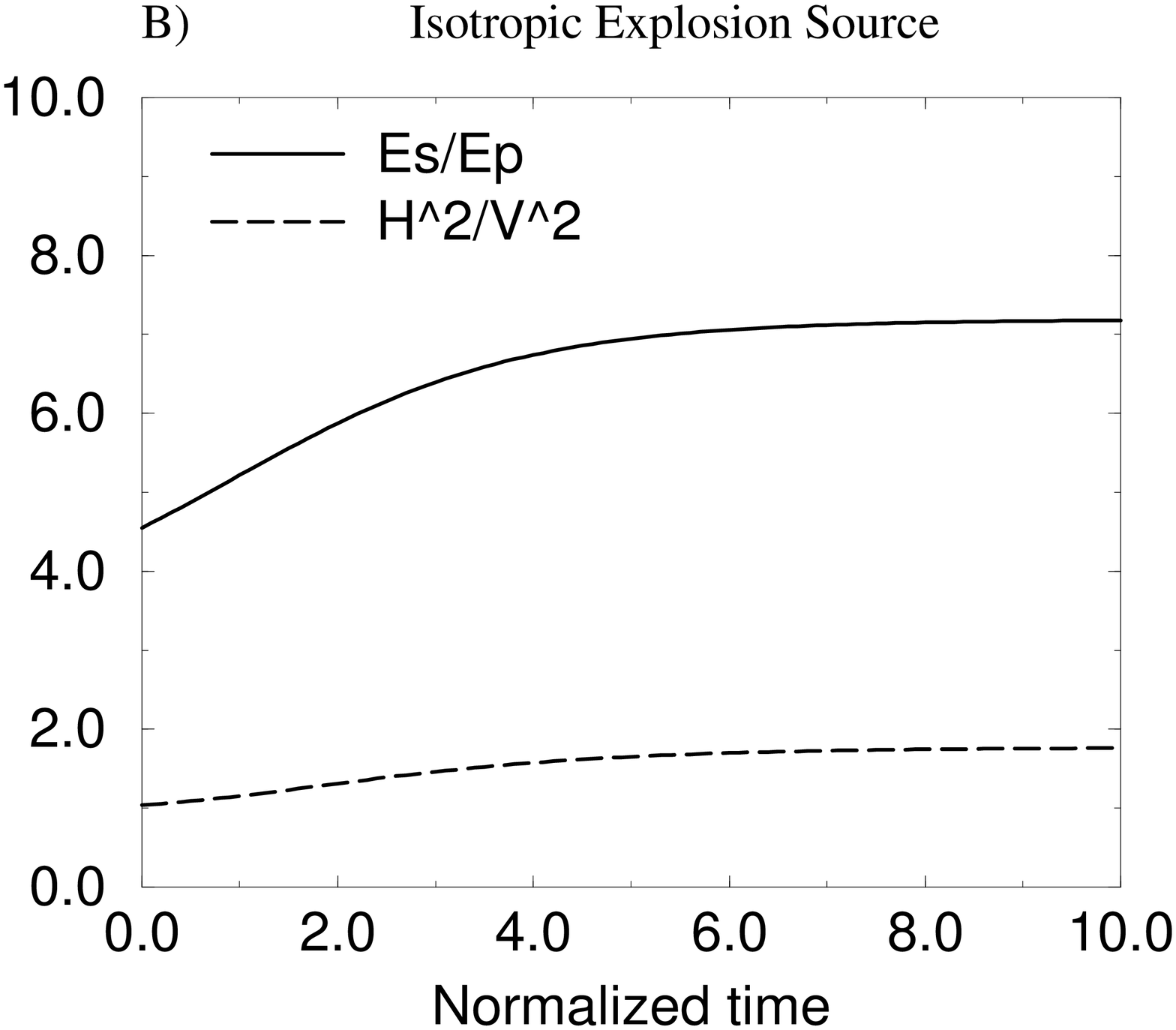,angle=-90,height=5cm,width=8cm}}
\end{minipage}
\begin{minipage}[t][5cm][s]{8cm}
{\psfig{figure=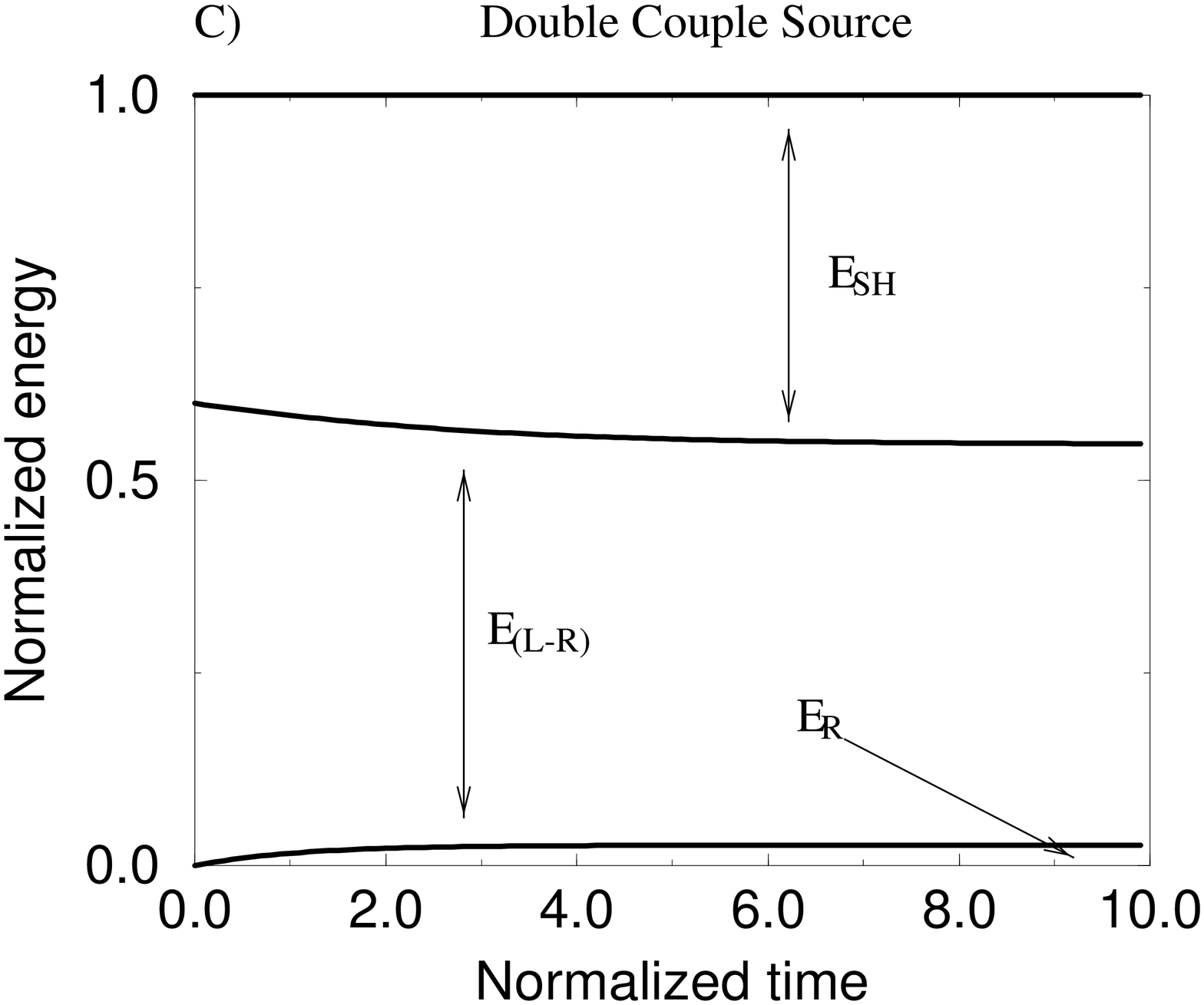,angle=-90,height=5cm,width=8cm}}
\end{minipage}\hfill
\begin{minipage}[t][5cm][s]{8cm}
{\psfig{figure=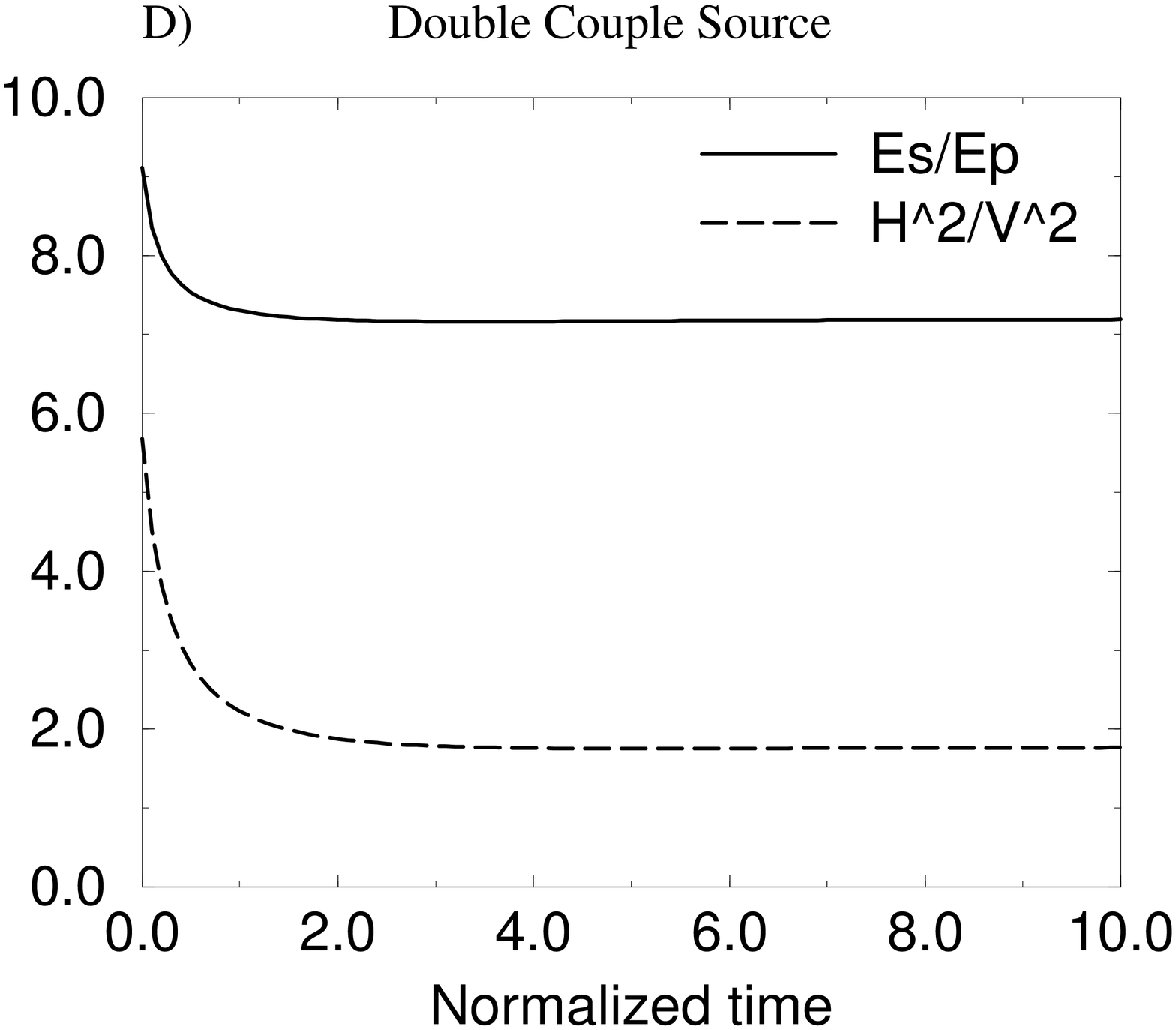,angle=-90,height=5cm,width=8cm}}
\end{minipage}
\end{figure}

\newpage

\begin{figure}
{\psfig{figure=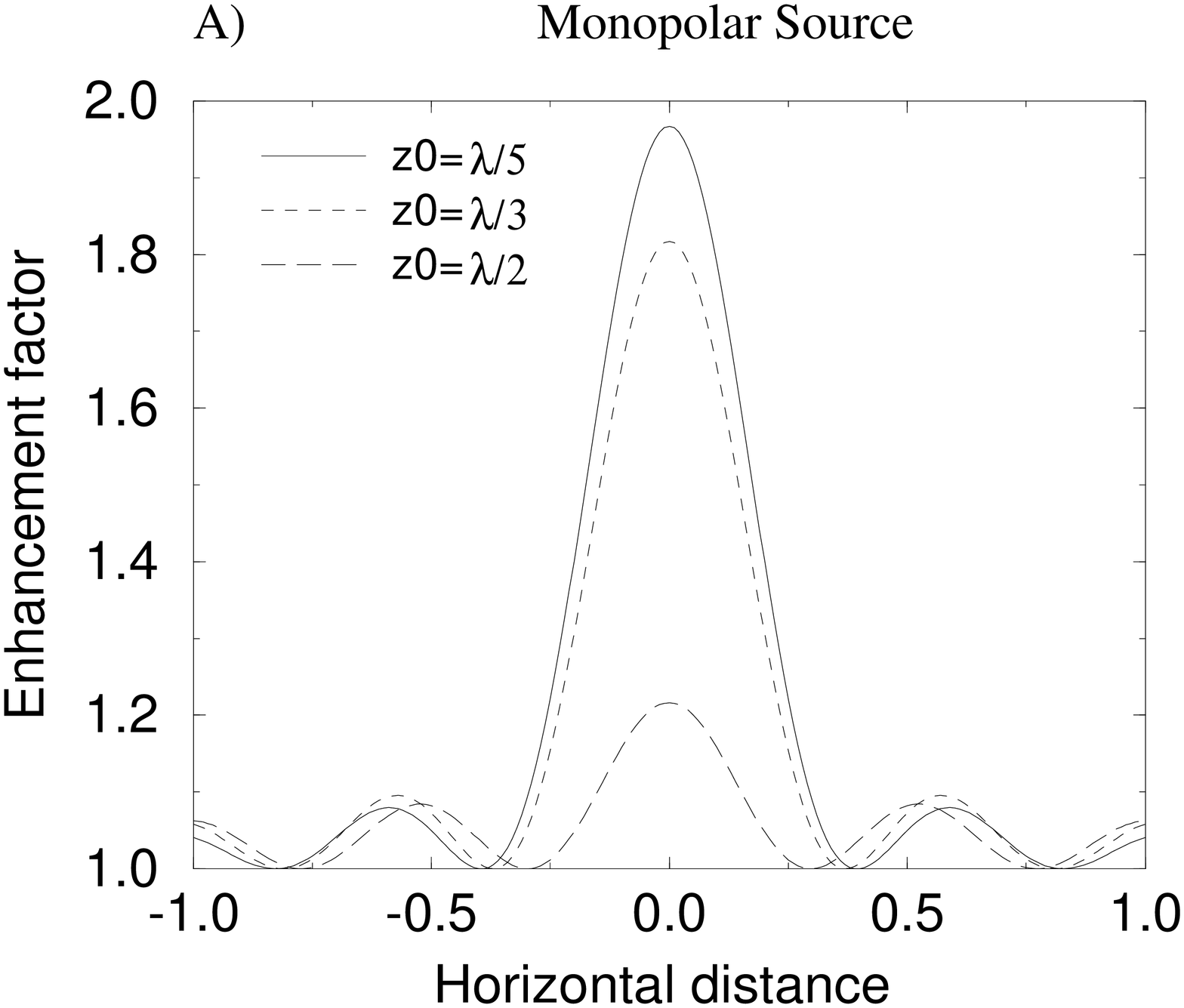,angle=-90,height=7cm,width=8.5cm}}
{\psfig{figure=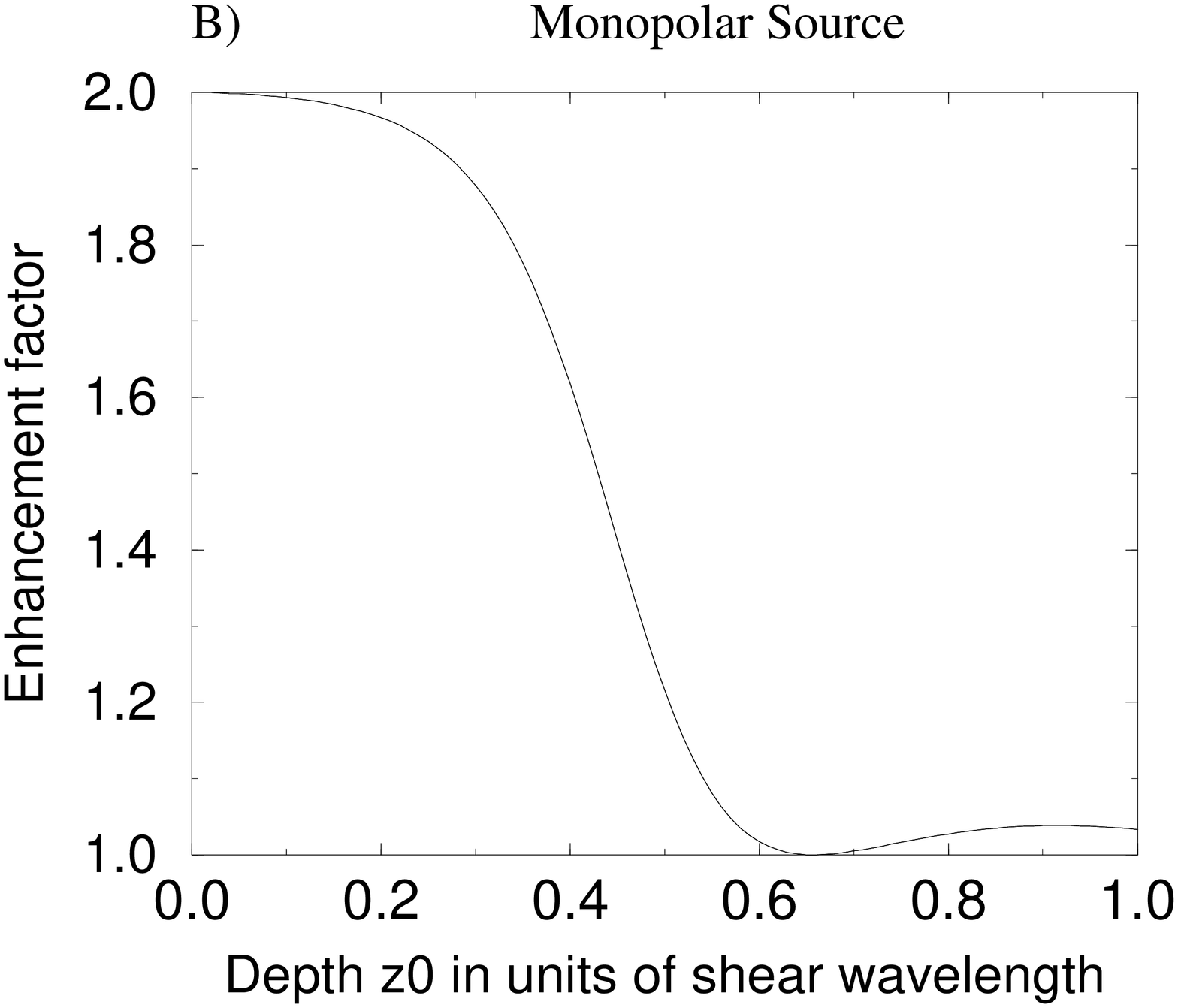,angle=-90,height=7cm,width=8.5cm}}
\end{figure}

\newpage

\begin{figure}
{\psfig{figure=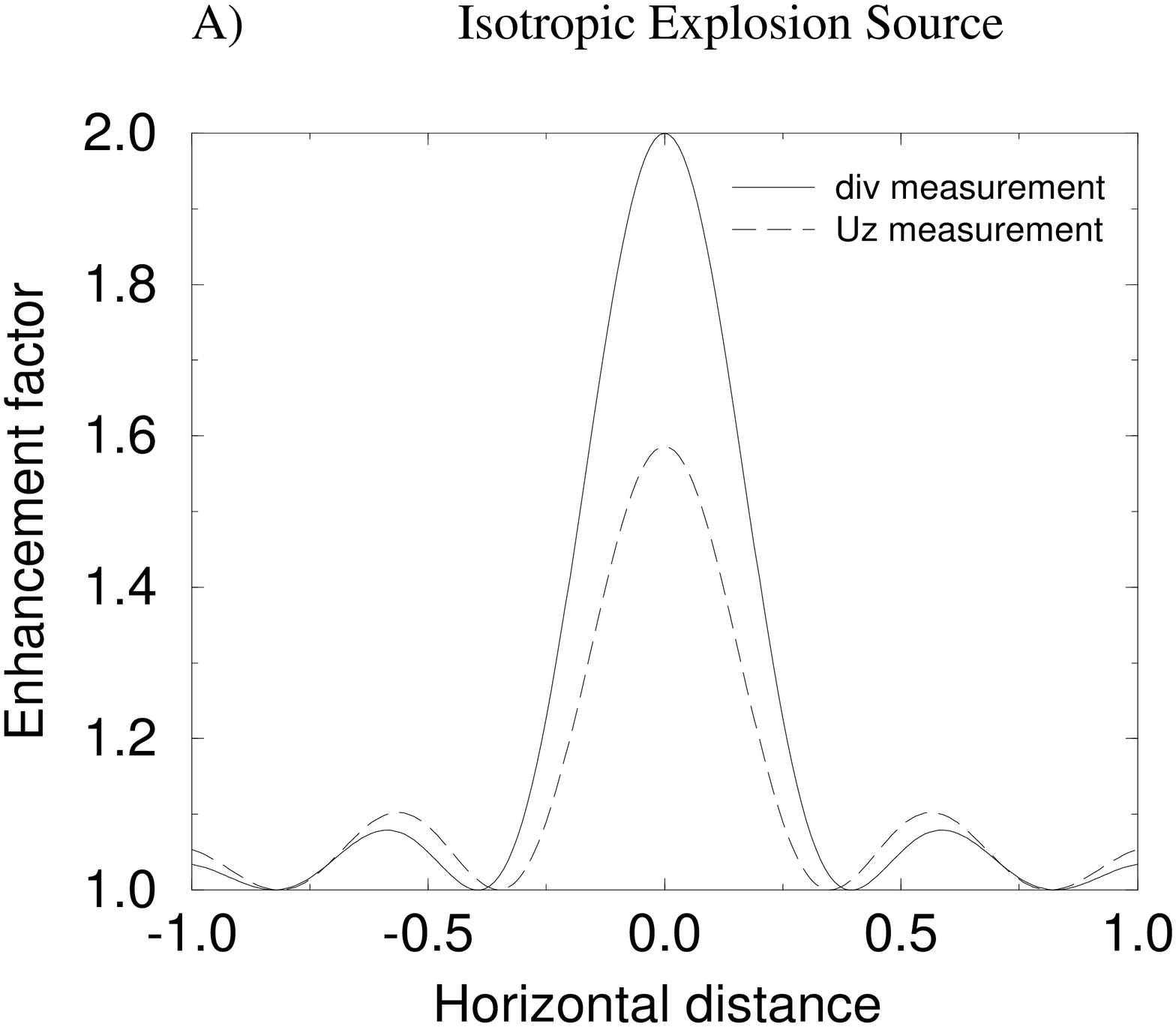,angle=-90,height=7cm,width=8.5cm}}
{\psfig{figure=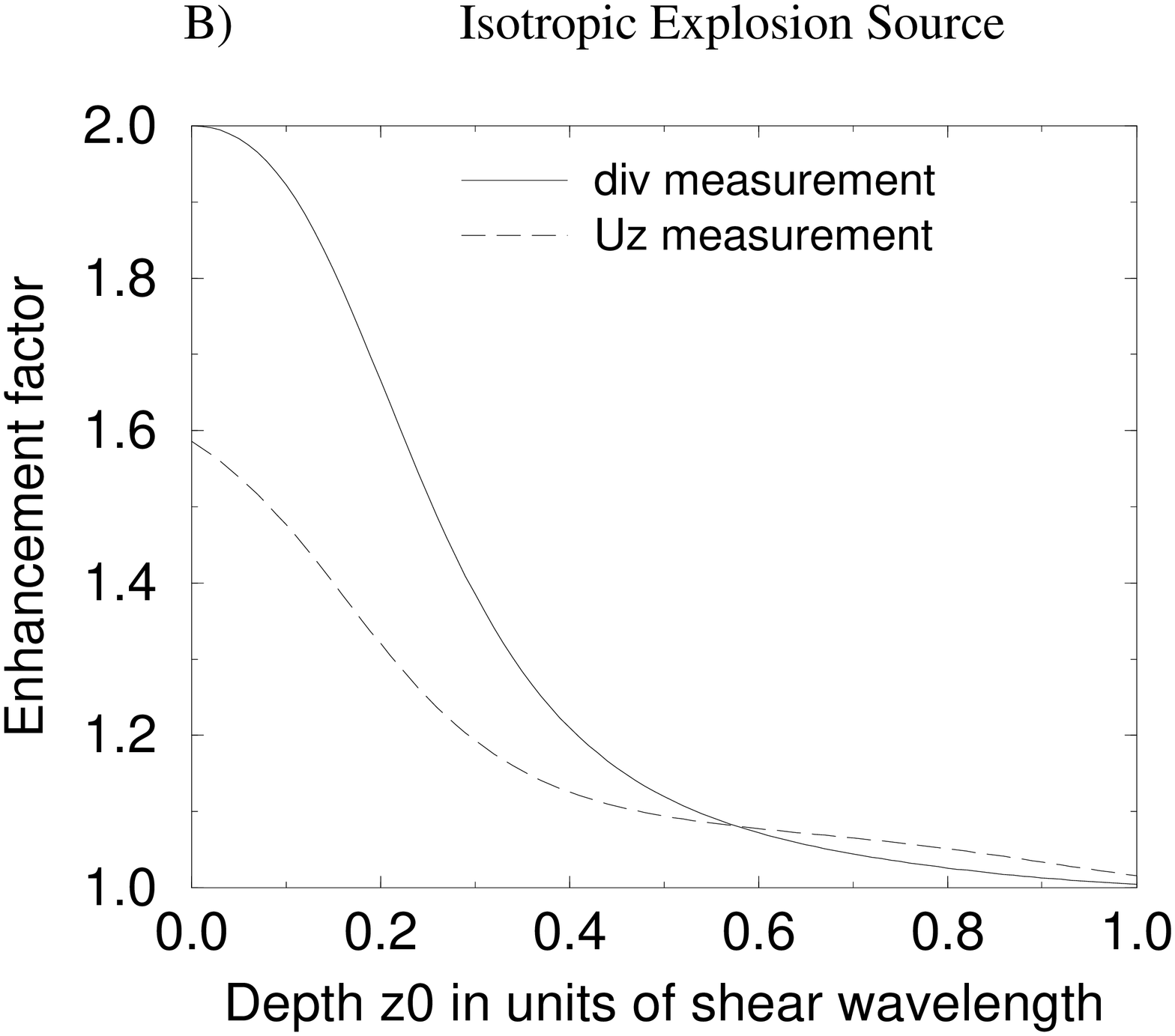,angle=-90,height=7cm,width=8.5cm}}
\end{figure}

\newpage

\begin{figure}
{\psfig{figure=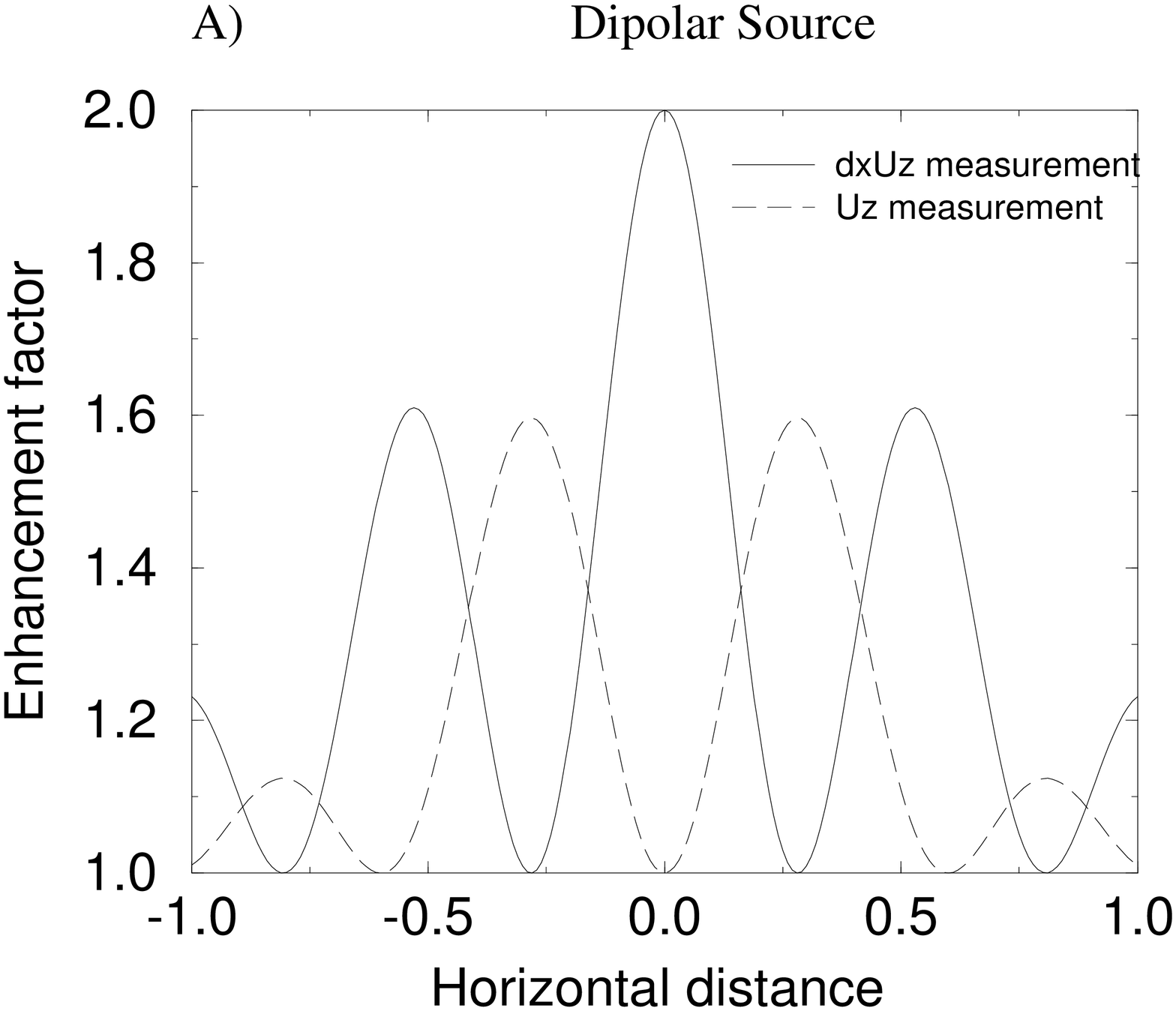,angle=-90,height=7cm,width=8.5cm}}
{\psfig{figure=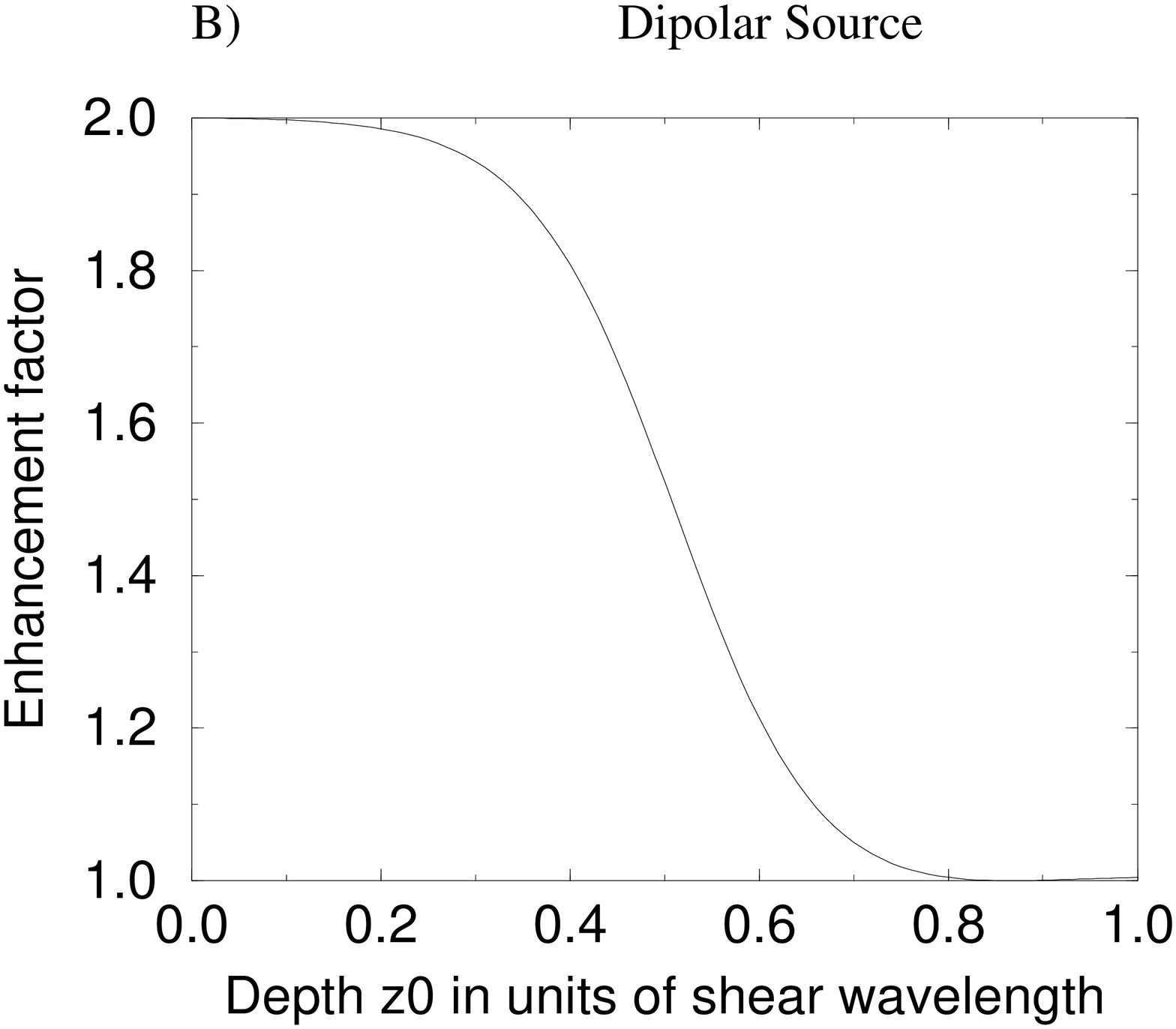,angle=-90,height=7cm,width=8.5cm}}
\end{figure}

\newpage

\begin{figure}
{\psfig{figure=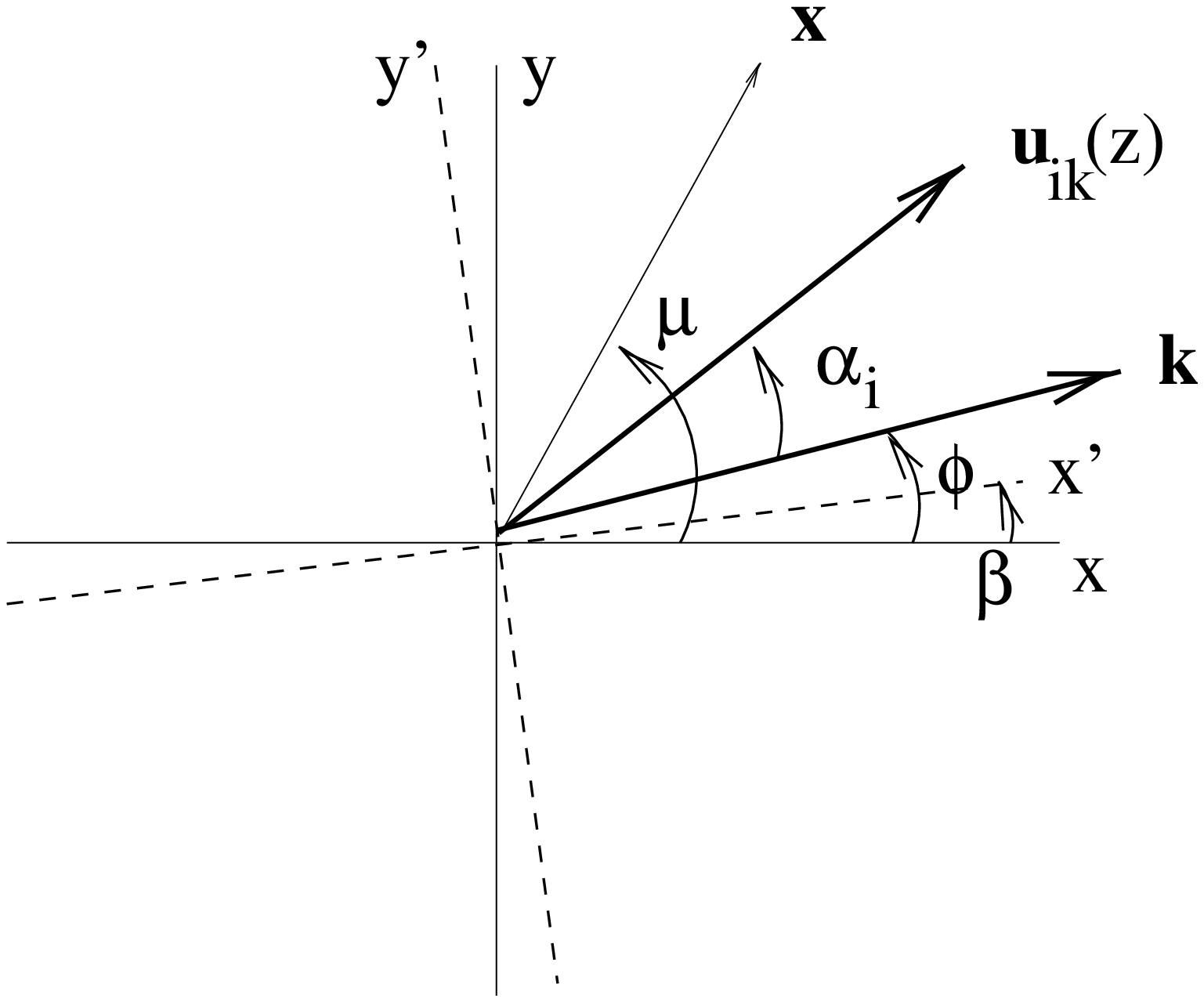,angle=-90,height=7.0cm,width=8.5cm}}
\end{figure}

\newpage

\begin{figure}
{\psfig{figure=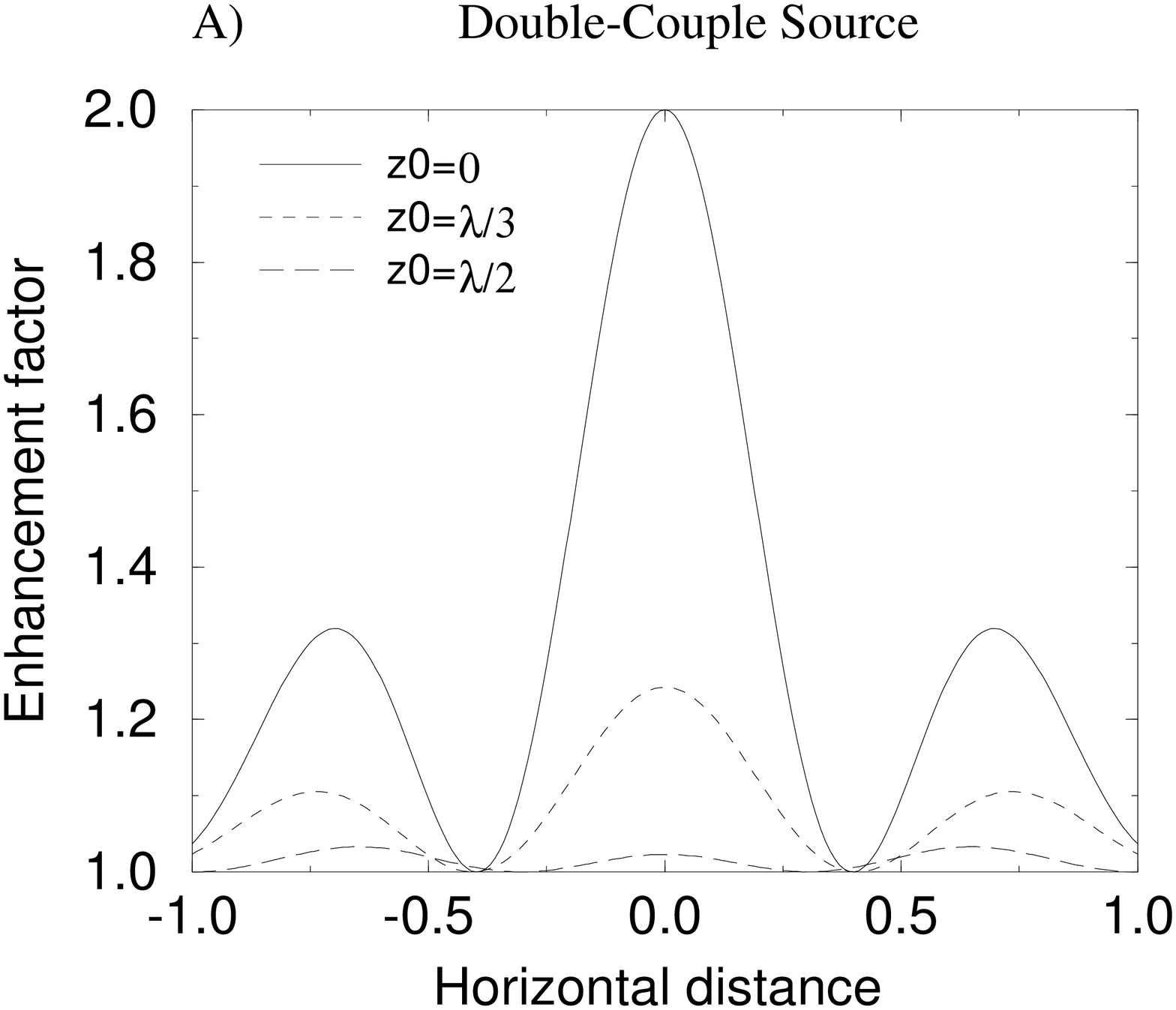,angle=-90,height=7cm,width=8.5cm}}
{\psfig{figure=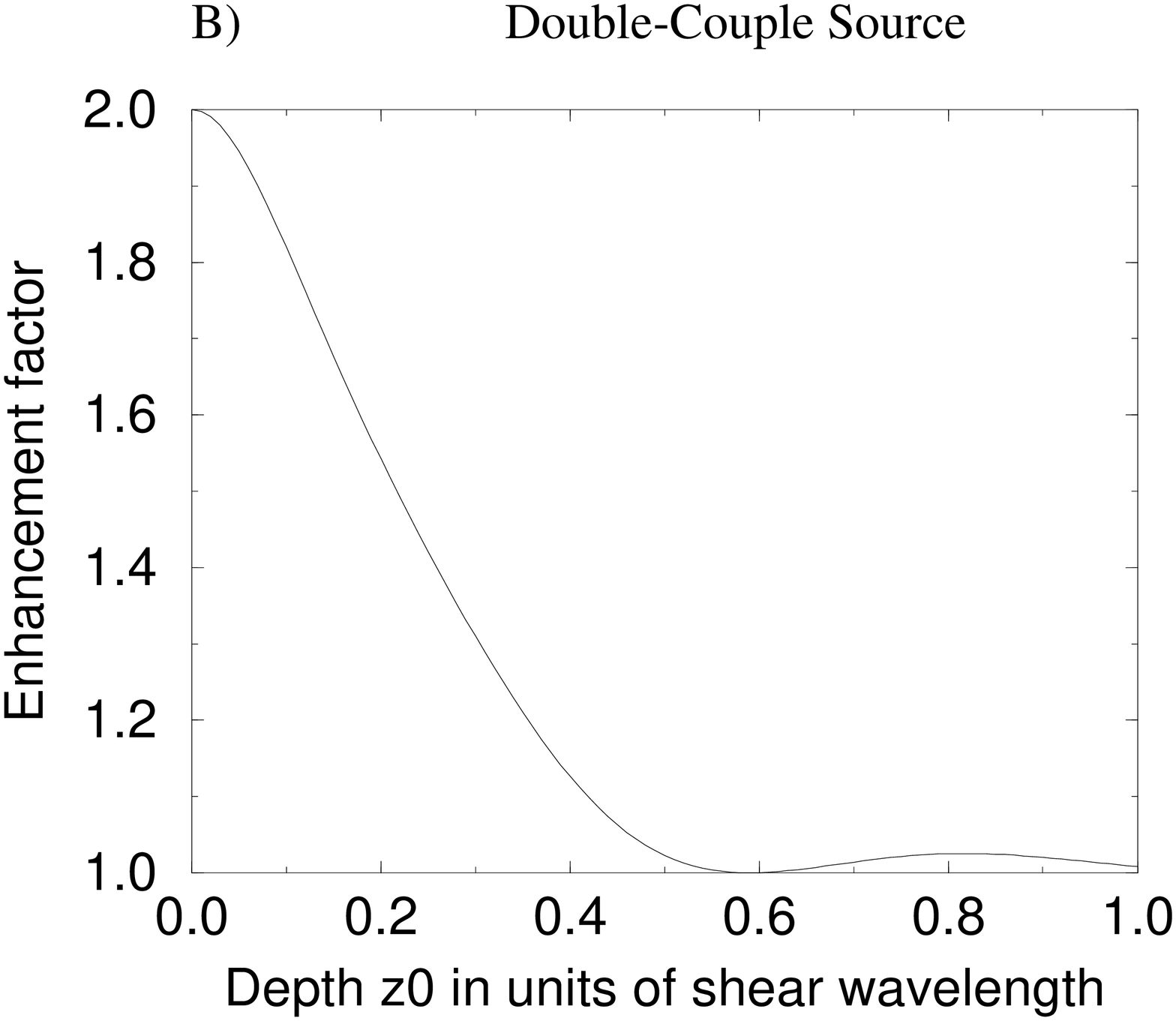,angle=-90,height=7cm,width=8.5cm}}
\end{figure}

\end{document}